\documentclass[useAMS, usenatbib, fleqn]{mn2e}

\usepackage{times}
\usepackage{url}
\usepackage{amsmath}
\usepackage{graphicx}
\usepackage{subfig}
\usepackage{xspace}
\usepackage{float}
\usepackage{caption}
\usepackage{amsfonts}
\usepackage{amssymb}
\usepackage{multirow}
\usepackage{color}
\usepackage[breaklinks, colorlinks, citecolor=blue, linkcolor=black, urlcolor=black]{hyperref}

\newcommand{\sqdeg}{{deg$^2$}\xspace}%
\newcommand{\healpix}{{\textsc HEALPix}\xspace}
\newcommand{\photoz}{photo-$z$\xspace}

\newcommand{\ie}{\textit{{\it i.e.},~}}
\newcommand{\eg}{\textit{{\it e.g.},~}}

\newcommand{\figref}[1]{Figure~\ref{#1}}

\newcommand{\nside}{{\rm N_{side}}}

\newcommand{\be}[1]{\begin{equation}}
\newcommand{\ee}[1]{\end{equation}}
\newcommand{\equ}[1]{\begin{equation}#1\end{equation}}

\newcommand{\bl}[1]{\textcolor{black}{#1}}

\pagerange{\pageref{firstpage}--\pageref{lastpage}} \pubyear{2015}
\def\LaTeX{L\kern-.36em\raise.3ex\hbox{a}\kern-.15em
    T\kern-.1667em\lower.7ex\hbox{E}\kern-.125emX}

\title[Observing conditions systematics in DES SV]
{Mapping and simulating systematics due to spatially-varying observing conditions in DES Science Verification data}

\author[Leistedt, Peiris, Elsner, Benoit-L\'{e}vy et al]{
\parbox{18cm}{
B.~Leistedt,$^{1}$\thanks{Contact email: boris.leistedt.11@ucl.ac.uk}
H.V.~Peiris,$^{1}$
F.~Elsner,$^{1}$
A.~Benoit-L{\'e}vy,$^{1}$
A.~Amara,$^{2}$
A.~H.~Bauer,$^{3}$
M.~R.~Becker,$^{4,5}$
C.~Bonnett,$^{6}$
C.~Bruderer,$^{2}$
M.~T.~Busha,$^{5,7}$
M.~Carrasco~Kind,$^{8,9}$
C.~Chang,$^{2}$
M.~Crocce,$^{3}$
L.~N.~da Costa,$^{10,11}$
E.~Gaztanaga,$^{3}$
E.~M.~Huff,$^{12,13}$
O.~Lahav,$^{1}$
A.~Palmese,$^{1}$
W.J.~Percival,$^{14}$
A.~Refregier,$^{2}$
A.~J.~Ross,$^{12}$
E.~Rozo,$^{15}$
E.~S.~Rykoff,$^{5,7}$
C.~S\'{a}nchez,$^{6}$
I.~Sadeh,$^{1}$
I.~Sevilla-Noarbe,$^{16,8}$
F.~Sobreira,$^{17,10}$
E.~Suchyta,$^{12,13}$
M.~E.~C.~Swanson,$^{9}$
R.~H.~Wechsler,$^{4,5,7}$
F.~B.~Abdalla,$^{1}$
S.~Allam,$^{17}$
M.~Banerji,$^{18,19}$
G.~M.~Bernstein,$^{20}$
R.~A.~Bernstein,$^{21}$
E.~Bertin,$^{22,23}$
S.~L.~Bridle,$^{24}$
D.~Brooks,$^{1}$
E.~Buckley-Geer,$^{17}$
D.~L.~Burke,$^{5,7}$
D.~Capozzi,$^{14}$
A.~Carnero~Rosell,$^{10,11}$
J.~Carretero,$^{3,6}$
C.~E.~Cunha,$^{5}$
C.~B.~D'Andrea,$^{14}$
D.~L.~DePoy,$^{25}$
S.~Desai,$^{26,27}$
H.~T.~Diehl,$^{17}$
P.~Doel,$^{1}$
T.~F.~Eifler,$^{20,28}$
A.~E.~Evrard,$^{29,30}$
A.~Fausti Neto,$^{10}$
B.~Flaugher,$^{17}$
P.~Fosalba,$^{3}$
J.~Frieman,$^{17,31}$
D.~W.~Gerdes,$^{30}$
D.~Gruen,$^{32,33}$
R.~A.~Gruendl,$^{8,9}$
G.~Gutierrez,$^{17}$
K.~Honscheid,$^{12,13}$
D.~J.~James,$^{34}$
M.~Jarvis,$^{20}$
S.~Kent,$^{17}$
K.~Kuehn,$^{35}$
N.~Kuropatkin,$^{17}$
T.~S.~Li,$^{25}$
M.~Lima,$^{36,10}$
M.~A.~G.~Maia,$^{10,11}$
M.~March,$^{20}$
J.~L.~Marshall,$^{25}$
P.~Martini,$^{12,37}$
P.~Melchior,$^{12,13}$
C.~J.~Miller,$^{29,30}$
R.~Miquel,$^{38,6}$
R.~C.~Nichol,$^{14}$
B.~Nord,$^{17}$
R.~Ogando,$^{10,11}$
A.~A.~Plazas,$^{28}$
K.~Reil,$^{7}$
A.~K.~Romer,$^{39}$
A.~Roodman,$^{5,7}$
E.~Sanchez,$^{16}$
B.~Santiago,$^{40,10}$
V.~Scarpine,$^{17}$
M.~Schubnell,$^{30}$
R.~C.~Smith,$^{34}$
M.~Soares-Santos,$^{17}$
G.~Tarle,$^{30}$
J.~Thaler,$^{41}$
D.~Thomas,$^{14, 43}$
V.~Vikram,$^{42}$
A.~R.~Walker,$^{34}$
W.~Wester,$^{17}$
Y.~Zhang,$^{30}$
J.~Zuntz$^{24}$
}\medskip\\
\vspace*{-6mm} Affiliations are listed at the end of the paper }

\begin{document}
\maketitle 
\begin{abstract}
Spatially-varying depth and characteristics of observing conditions, such as seeing, airmass, or {sky background}, are major sources of systematic uncertainties in modern galaxy survey analyses, in particular in deep multi-epoch surveys. We present a framework to extract and project these sources of systematics onto the sky, and apply it to the Dark Energy Survey (DES) to map the observing conditions of the Science Verification (SV) data. The resulting distributions and maps of  sources of systematics are used in several analyses of DES SV to perform detailed null tests with the data, and also to incorporate systematics in survey simulations. We illustrate the complementarity of these two approaches by comparing the SV data with the BCC-UFig, a synthetic sky catalogue generated by forward-modelling of the DES SV images. We analyse the BCC-UFig simulation to construct galaxy samples mimicking those used in SV galaxy clustering studies. We show that the spatially-varying survey depth imprinted in the observed galaxy densities and the redshift distributions of the SV data are successfully reproduced by the simulation and well-captured by the maps of observing conditions. The combined use of the maps, the SV data and the BCC-UFig simulation allows us to quantify the impact of spatial systematics on $N(z)$, the redshift distributions inferred using photometric redshifts. We conclude that spatial systematics in the SV data are mainly due to seeing fluctuations and are under control in current clustering and weak lensing analyses. However, they will need to be carefully characterised in upcoming phases of DES in order to avoid biasing the inferred cosmological results. The framework presented here is relevant to all multi-epoch surveys, and will be essential for exploiting future surveys such as the Large Synoptic Survey Telescope, which will require detailed null-tests and realistic end-to-end image simulations to correctly interpret the deep, high-cadence  observations of the sky. 
\end{abstract}

\section{Introduction}

The Dark Energy Survey \citep[DES,][]{2005astro.ph.10346T} began in 2012 and will observe during at least five seasons to cover $\sim$ 5000 square degrees of the Southern sky, in five optical bands ({\it grizY}). When completed, DES will cover a volume of the Universe up to 20 times greater than the Sloan Digital Sky Survey (SDSS, \citealt{Gunn2006}), the largest optical survey to date. Hence, DES will provide an enormous legacy data set useful in a range of astrophysical and cosmological studies. It is thus essential to develop approaches to robustly analyse DES data while accounting for statistical and systematic uncertainties.

The primary science goal of DES is to {uncover} the nature of dark energy using a combination of cosmological observables. In addition to expansion rate measurements using supernova light curves, DES will rely on probes of the growth rate such as the clustering and gravitational lensing of galaxies and clusters of galaxies. Exploiting these observables to probe dark energy requires exquisite control over the spatial coverage and calibration of the survey. Spatial fluctuations in the depth or quality of the data (\eg the properties of the sky noise, the photometry, or galaxy ellipticity measurements) can impact the galaxy catalogues and lead to systematic biases in cosmological analyses. All ongoing and future surveys will be limited by our ability to identify and mitigate such systematics.  

Establishing an exhaustive list of the sources of potential systematics in cosmological measurements is beyond the scope of this paper. However, it is worth recalling that systematics in clustering and cosmic shear studies are mostly rooted in astrophysical foregrounds (extinction by dust or obscuration by bright stars), observing conditions (\eg seeing, sky noise, airmass), or processing and calibration (such as the quality of the photometry or the point spread function). These affect the probability of detecting sources and also their properties, yielding non-trivial distortions in the reduced data, in particular the galaxy catalogues. In DES, various efforts are dedicated to modelling or capturing the complicated \textit{transfer function} connecting the raw data to the final galaxy catalogues. {For instance, the Ultra Fast Image simulator \citep[UFig, ][]{Berge2013ufig} is used to create simulated DES images, which are then processed in a similar manner to the real  data. This approach has been investigated \eg to characterise systematics in shear measurements \citep{2015arXiv150402778B}. UFig was also interfaced with the BCC N-body simulations \citep{busha2013bcc} by \cite{Chang2014bccufig} in order to forward-model the survey transfer function with known underlying astrophysics and cosmology. In this paper, we test this transfer function and investigate how well the BCC-UFig is able to reproduce physical characteristics (\eg redshift distributions) and systematics (\eg spurious galaxy density fluctuations) found in the DES Science Verification (SV) data.} By contrast, \textsc{Balrog}\footnote{https://github.com/emhuff/Balrog} (\citealt{2015Balrog}, used in \citealt{Melchior:2014yap}) takes the approach of populating real DES images with simulated galaxies in order to measure the effective transfer function of the survey. These complementary efforts will be improved in the coming years to fully exploit DES data. 

The observing conditions and astrophysical foregrounds unavoidably vary across the survey footprint (\eg  nightly variations of seeing, or colour reddening by Galactic dust). This paper focuses on mapping these sources of systematics onto the sky. This operation is analogous to the construction of foreground templates for the analysis of cosmic microwave background (CMB) data \cite[\eg][]{Teg97, SlosarSeljak2004modeproj, Planck2013component}. {Such templates are used in numerous analyses of single-epoch surveys like SDSS, in particular galaxy and quasar clustering measurements \citep[\eg][]{THS1998future,  2002ApJ...579...48S, ross2012systematics, ho2012cosmoweights, Leistedt2013excessdr6, Leistedt:2014wia, agarwalho2013sys}, and are being used in analyses of SV data \citep[\eg][]{2015arXiv150403002V, Crocce2015dessvclustering, Jarvis2015svshear, Becker2015svshear, Giannantonio2015cmbxlssdessv}}. More generally, templates of potential sources of systematics can be used to carry out null tests with the data or model their contamination. As detailed below, multi-epoch surveys such as DES require a dedicated projection framework. In addition, the extracted observing conditions can be incorporated in image simulations to mimic the survey properties.

This paper is organised as follows. In Section 2 we present a scheme to map multi-epoch survey data onto the sky, and apply it to DES {SV} data. We present and analyse the resulting maps of sources of observational systematics. In Section 3 we use these maps to analyse the SV data and the BCC-UFig simulations, and show the impact of observational systematics on the measured galaxy densities and on the redshift distributions inferred using photometric redshifts. In Section 4 we conclude and discuss the impact and future extensions of this work.

\section{Mapping the properties of DES-SV images}

\subsection{Geometrical projection}

\begin{figure}
\hspace{-5mm}\begin{minipage}{9.4cm}\centering
\hspace{-3mm}\includegraphics[width=7.8cm, trim = 0.4cm 0.2cm 0cm 0.4cm, clip]{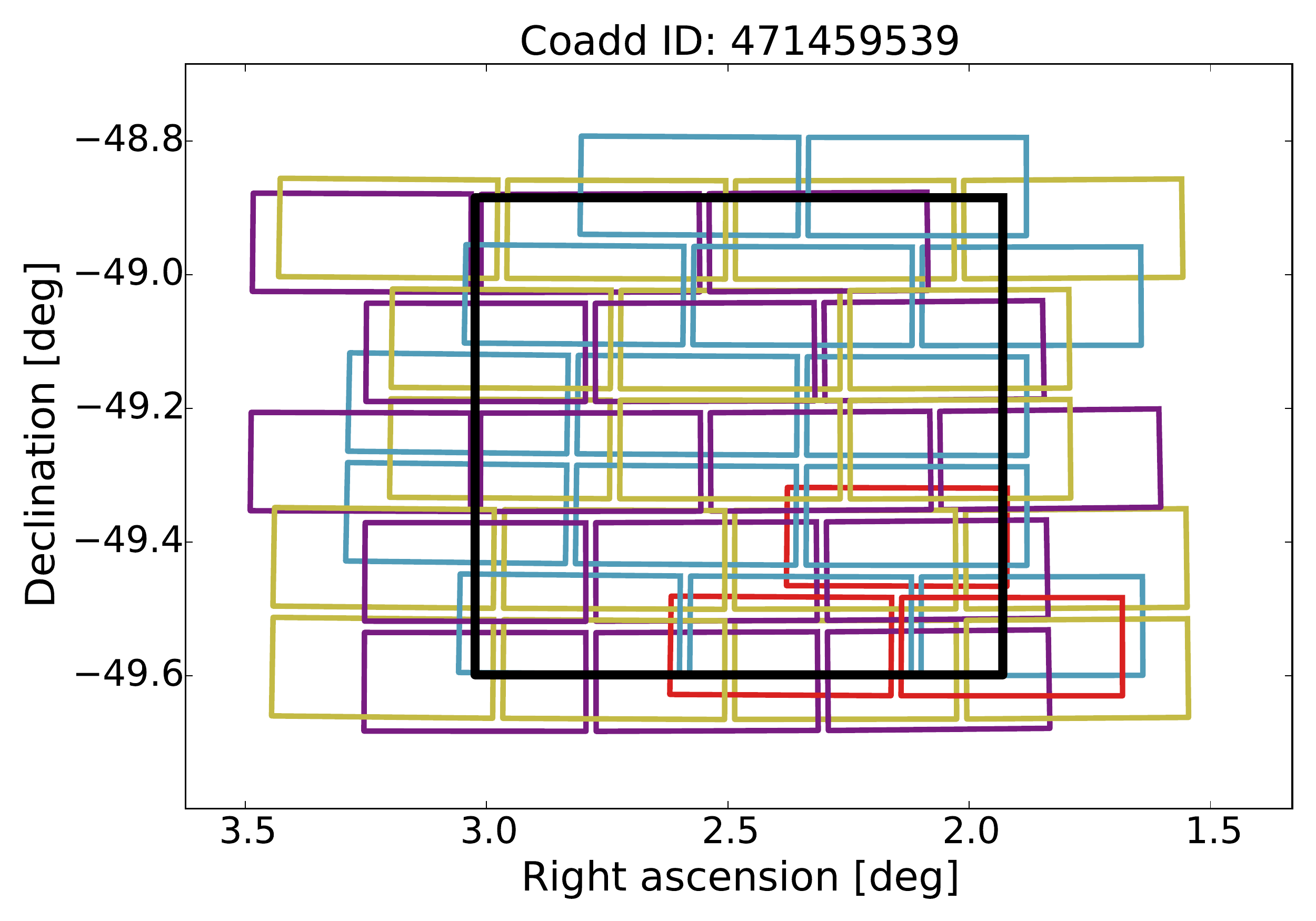}\\
\includegraphics[width=8.6cm, trim = 0cm 0.2cm 0cm 0cm, clip]{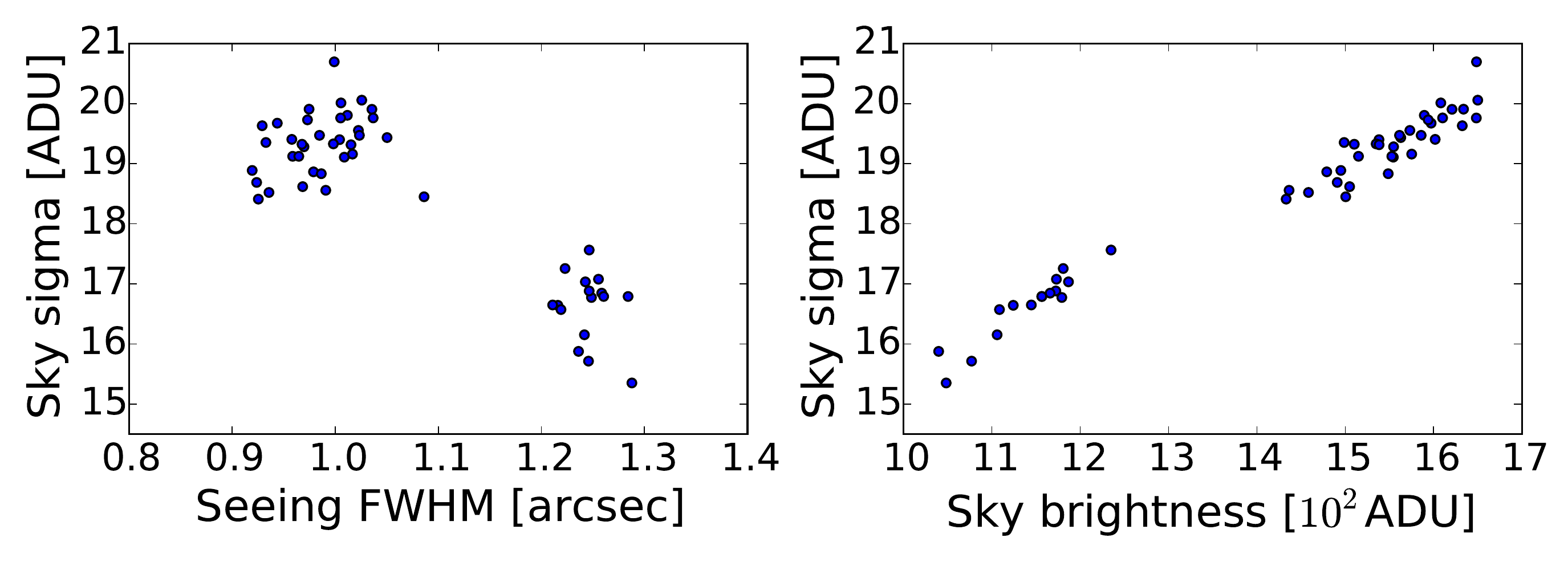}
\end{minipage}
\caption{\bl{Upper panel: geometrical projection of the single-exposure images coadded in an arbitrary tile of the DES Science Verification data (black contour). The colours correspond to different single-epoch pointings, with the relevant CCDs shown as individual rectangles. Lower panels: properties of the same set of CCDs, exhibiting significant variations and correlations. The nontrivial, spatially-varying geometrical overlap and image properties will result in spatially-varying systematics when analysing the galaxy catalogues.}}
\label{fig:ccdplot}
\end{figure}

Mapping potential sources of systematics, such as observing conditions, is a routine operation in modern galaxy surveys. For the SDSS, this mapping was relatively straightforward since SDSS was a single-epoch survey. Therefore, a direct mapping between sky position and images could be established\footnote{With the exception of the Stripe 82 region, the deeper multi-epoch programme of SDSS, and the small zones of overlap between the single-epoch images.} \citep[\eg][]{ross2011weights, ross2012systematics, Leistedt:2014wia}. In other words, any of the properties of SDSS images (\eg seeing) directly project onto the sky. This is no longer the case for DES, which is a multi-epoch survey where several single-epoch CCD images are processed and stacked into `coadd' images, from which galaxies and stars are then extracted. The nominal depth in the main DES survey requires up to ten tilings in each band, while deeper regions require an order of magnitude more (\ie in the SN fields, which are dedicated to the DES supernova programme). The coadding process is done in non-overlapping regions called `tiles', which are {$0.75\times0.75$} \sqdeg squares constructed to cover and uniquely decompose the entire DES footprint. As a consequence of the multi-epoch nature of DES, there is not a unique value of \eg seeing at each sky position, but rather a distribution of values corresponding to the coadded single-epoch images. This is illustrated in \figref{fig:ccdplot}, which shows the footprints and properties of a set of single-epoch images used in an arbitrary coadd, part of the DES-SV data (described in the next section). The seeing, airmass and  background noise (as well as many other properties not shown here) exhibit strong fluctuations and correlations. Combined with the non-trivial geometrical overlap between these images, this demonstrates the need for a flexible projection framework. In the standard processing pipeline, these images are processed and coadded in tiles (black line of \figref{fig:ccdplot}) with the DESDM software, using the software packages described in \cite{2011arXiv1109.6741S, 2012ApJ...757...83D, 2012SPIE.8451E..0DM}\footnote{Including SCAMP (astrometry, \citealt{2006ASPC..351..112B}), SWARP (image coaddition, \citealt{2002ASPC..281..228B}), PSFEx (modelling of the point- spread-function, \citealt{2011ASPC..442..435B}) and SExtractor (object detection and measurement, \citealt{1996A&AS..117..393B}).}. The operations performed in these codes unavoidably mix the image properties across the coadds and affect the properties of detected sources. The geometry of the DECam focal plane --- a hexagonal shape, with 62 science CCDs \citep{2015arXiv150402900F, 2008arXiv0810.3600H} --- may also be imprinted in the reduced data. Therefore, one would like to access the full distribution of the single-epoch properties, and connect it to the coadds, catalogues, and sky coordinates.

\begin{figure}
\centering
\includegraphics[width=3.6cm, trim = 4.7cm 2.8cm 3.0cm 12.8cm, clip]{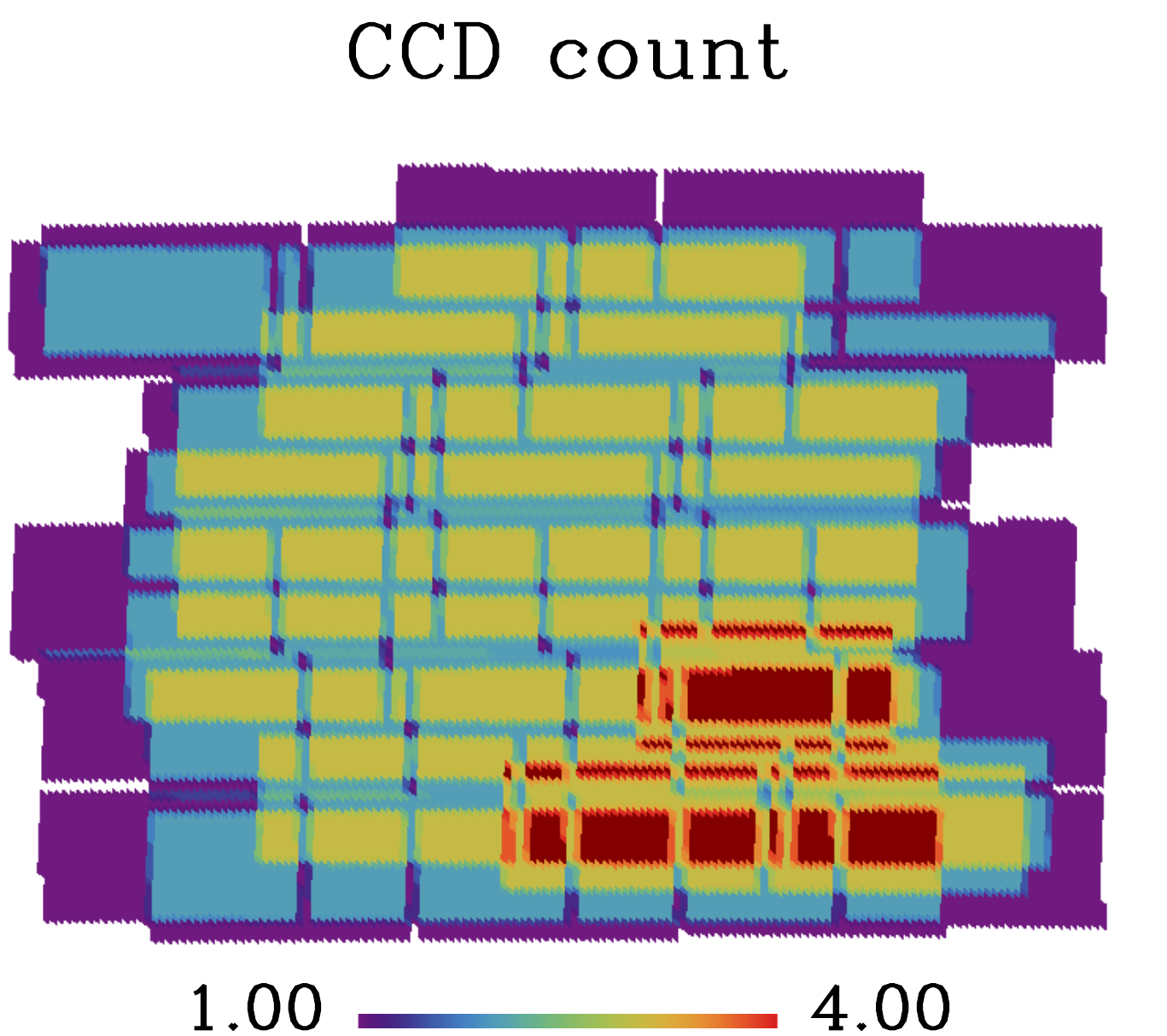} \quad
\includegraphics[width=3.6cm, trim = 4.7cm 2.8cm 3.0cm 12.8cm, clip]{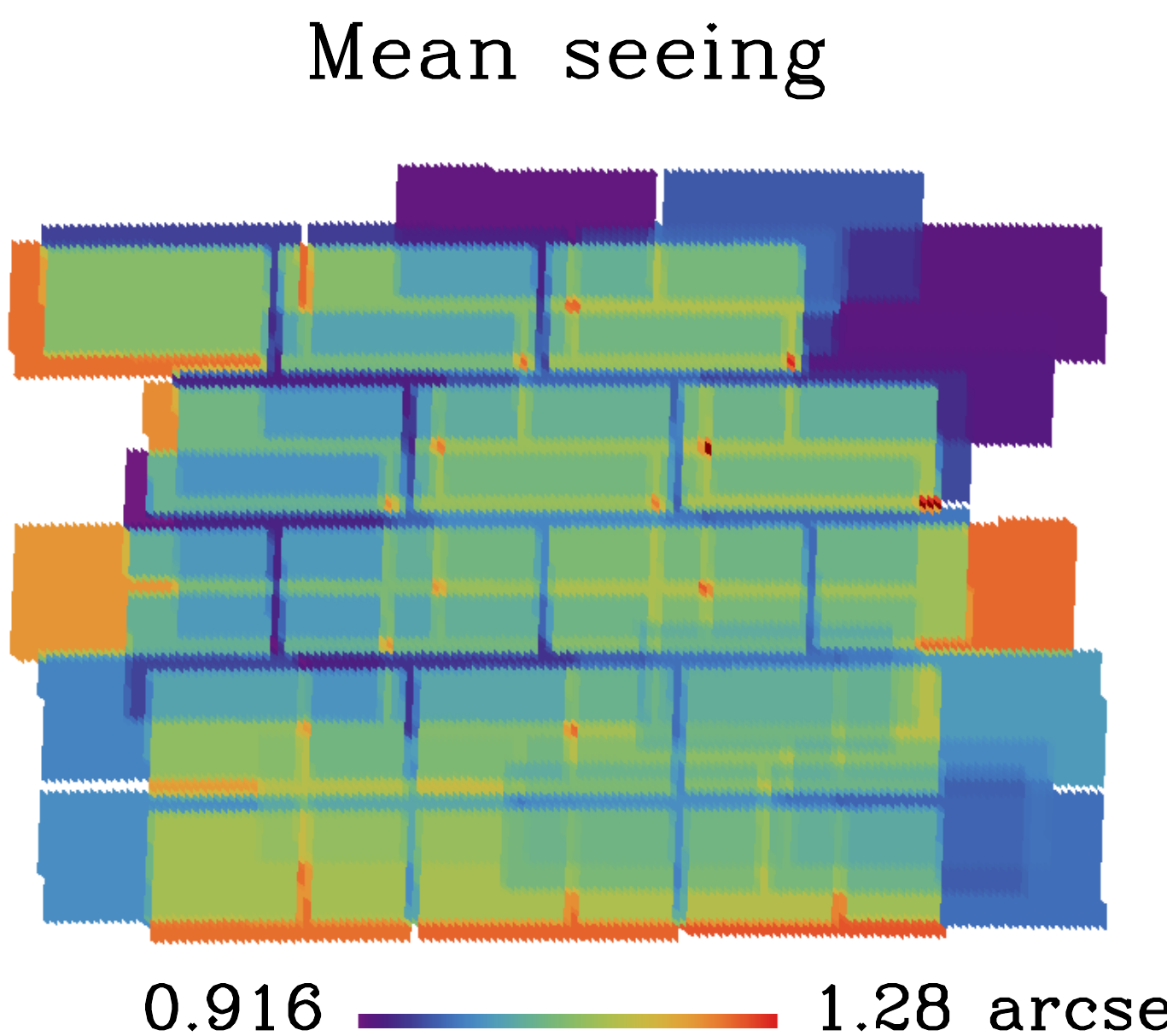} \vspace*{2mm}\\
\includegraphics[width=3.6cm, trim = 4.7cm 2.8cm 3.0cm 12.8cm, clip]{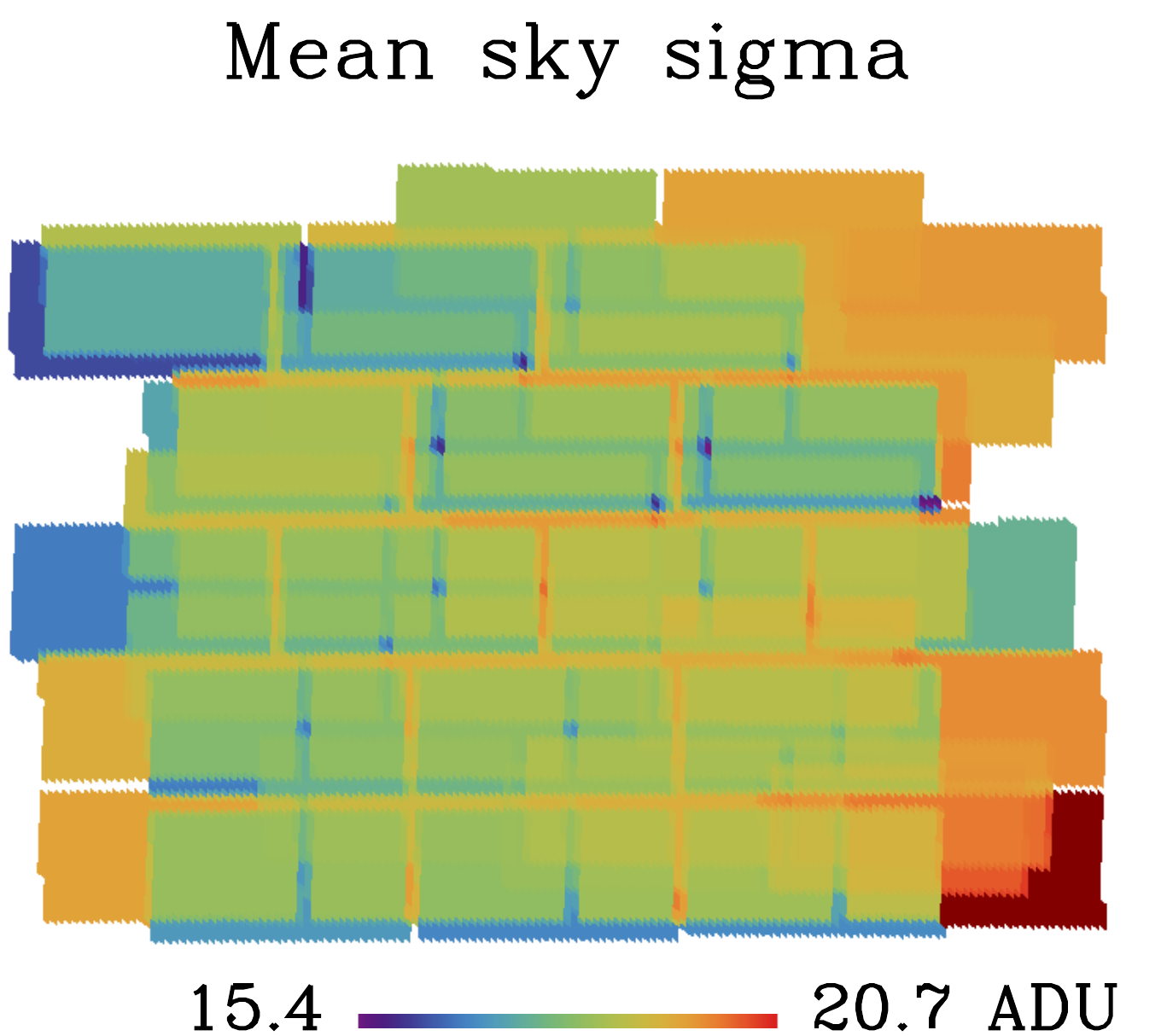} \quad
\includegraphics[width=3.6cm, trim = 4.7cm 2.8cm 3.0cm 12.8cm, clip]{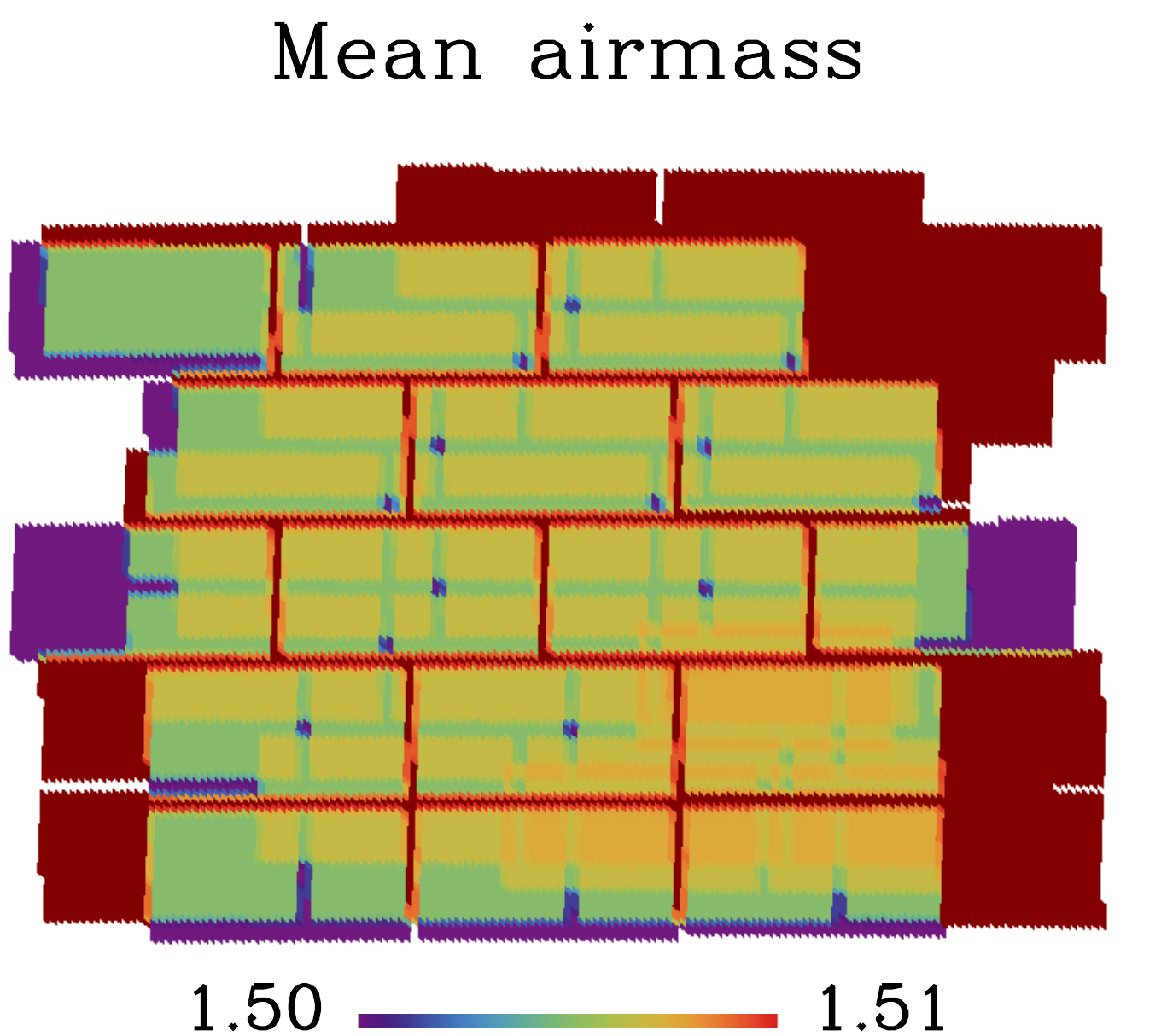}
\caption{Projection of the properties of the single-epoch images of \figref{fig:ccdplot}, showing how the time fluctuations and correlations are converted into spatial fluctuations. \bl{ADUs are Analog-Digital Units. }}
\label{fig:projccdplot}
\end{figure}

To construct sky maps of the single-epoch properties, we proceed as follows. We first {connect} the single-epochs and coadds, and keep track of which images were processed and coadded by the DESDM software. We then resolve the geometry of all images so that a given position on the sky is connected to a single coadd image and to a set of single-exposure CCDs. This is realised by accessing the images individually and using the WCS\footnote{WCS refers to the World Coordinate System of the FITS format \citep{2002A&A...395.1077C}.} transformations to convert local image coordinates into equatorial coordinates. We also make sure these transformations match the procedures used in the DES software\footnote{\bl{In particular, DES images make use of the WCS TPZ projection, built on the standard TAN projection and adding general polynomial corrections.}}. Finally, we employ the \healpix pixelisation \citep{healpix1} and connect the tree of geometrically-resolved images to \healpix pixels on the sky. 

The previous construction gives access to the full joint distribution of single-epoch and coadd image properties on the sky. This is a complicated object since each \healpix pixel contains a vector of image properties. As mentioned before, a crucial product is the projection of this joint distribution into scalar sky maps. This requires the computation of one value (such as a summary statistic) per pixel, \eg compressing the vector of seeing values in each pixel into mean, median, standard deviation, or even minimum and maximum values. This process can be done for any quantity of interest, with arbitrary weights. This is how any potential source of spatial systematics arising from single-epoch images can be mapped onto the sky. \figref{fig:projccdplot} shows the result of projecting some of the properties of the images of \figref{fig:ccdplot}. We see that the geometry of the CCDs as well as the relative orientations of the focal planes for the various exposures strongly affect the coverage and mean properties of the survey.

\subsection{Application to DES SV data}

\begin{figure}
\includegraphics[width=8.5cm, trim = 0.4cm 0.4cm 0.1cm 0.4cm, clip]{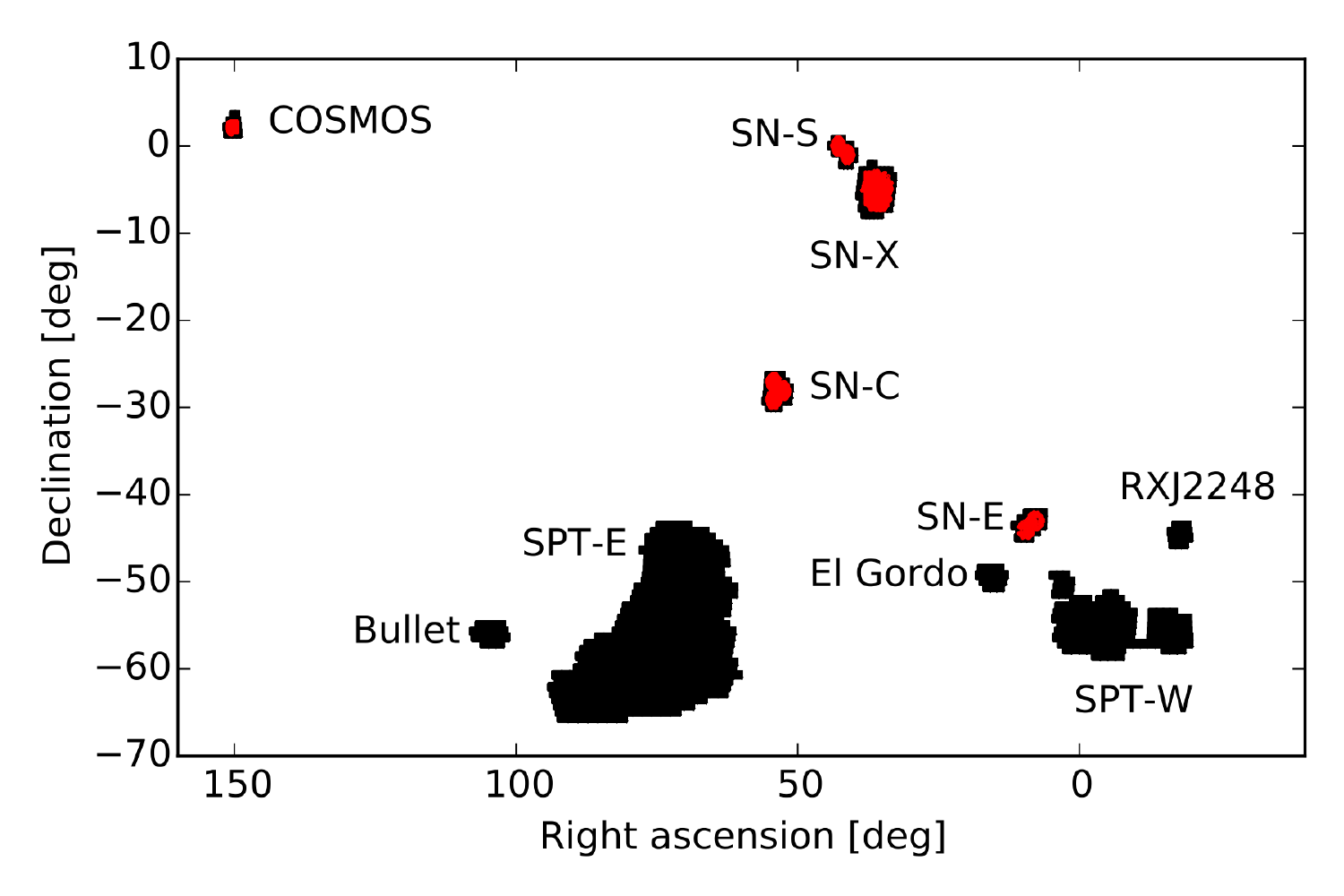}
\caption{The DES SV footprint, partitioned into several discontinuous regions, the largest being the SPT-E and W fields ($\approx 200$ and $50$ \sqdeg, respectively). The small red regions contain objects where spectroscopic redshifts are available, used to train the photometric redshift estimation codes, as discussed in Section 3.}
\label{fig:sva1_tiles}
\end{figure}

Science Verification (SV) data refers to the {testing data} acquired between November 2012 and February 2013, processed by the ``SVA1'' version of the DESDM pipeline \citep{Yanny2015} {and consisting of} 858 coadd tiles, 665 of which have data in all five {\it grizY} bands. The SV data cover more than 300 \sqdeg in total, split into contiguous regions of interest: the large SPT-E and SPT-W regions ($\approx 200$ and $50$ \sqdeg, respectively), the RXC J2248, Bullet, and El Gordo known rich clusters ($\approx 10$ \sqdeg each), COSMOS ($\approx 6$ \sqdeg) and the Supernovae fields SN-E, SN-X, SN-S, SN-C ($\approx 10$ \sqdeg each). The footprint of DES-SV is shown in \figref{fig:sva1_tiles} with the various fields labelled. 

\begin{figure}\centering
\includegraphics[width=4cm, trim = 2.9cm 3.3cm 1.3cm 2.5cm, clip]{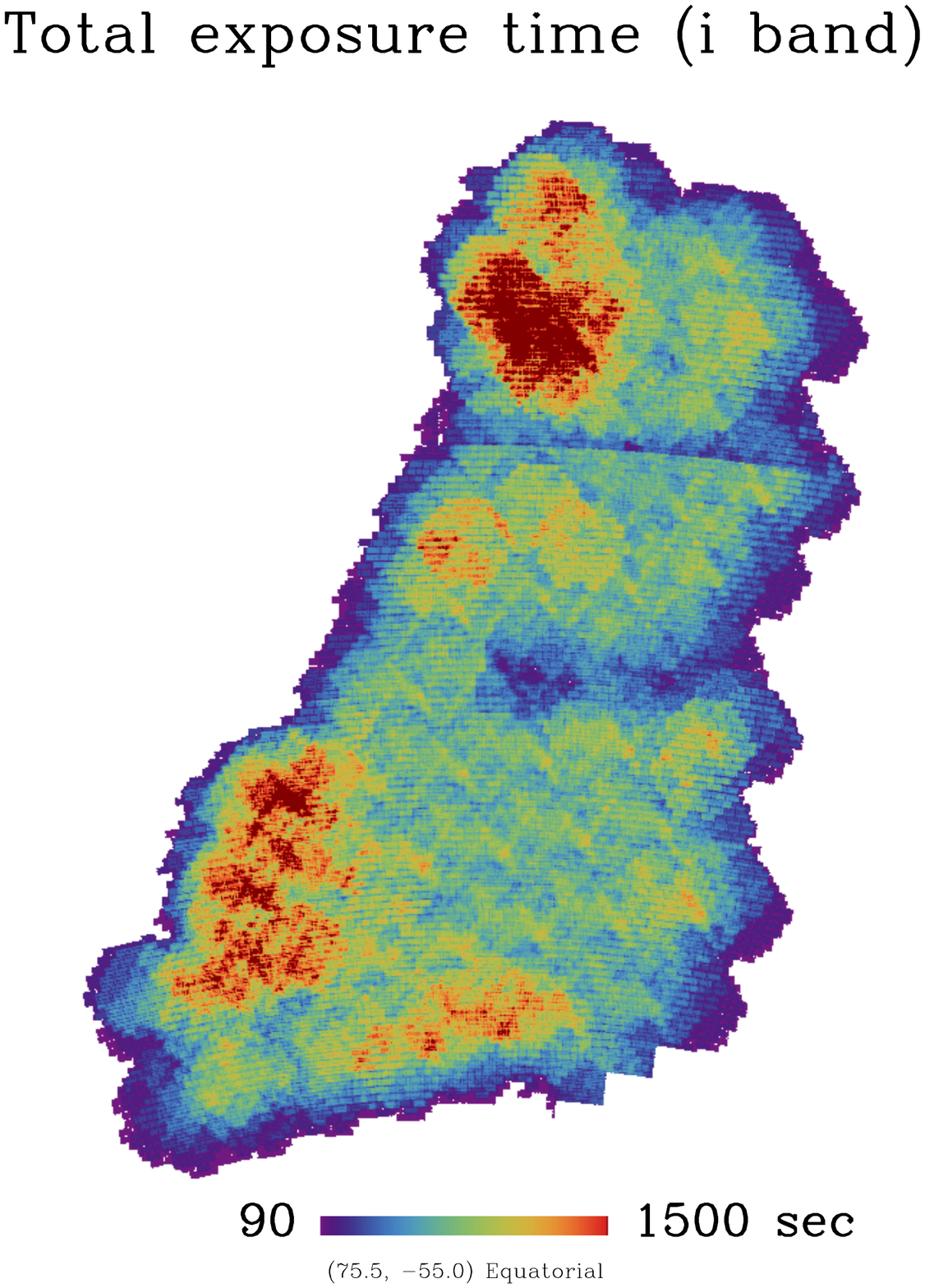}
\includegraphics[width=4cm, trim = 2.9cm 3.3cm 1.3cm 2.5cm, clip]{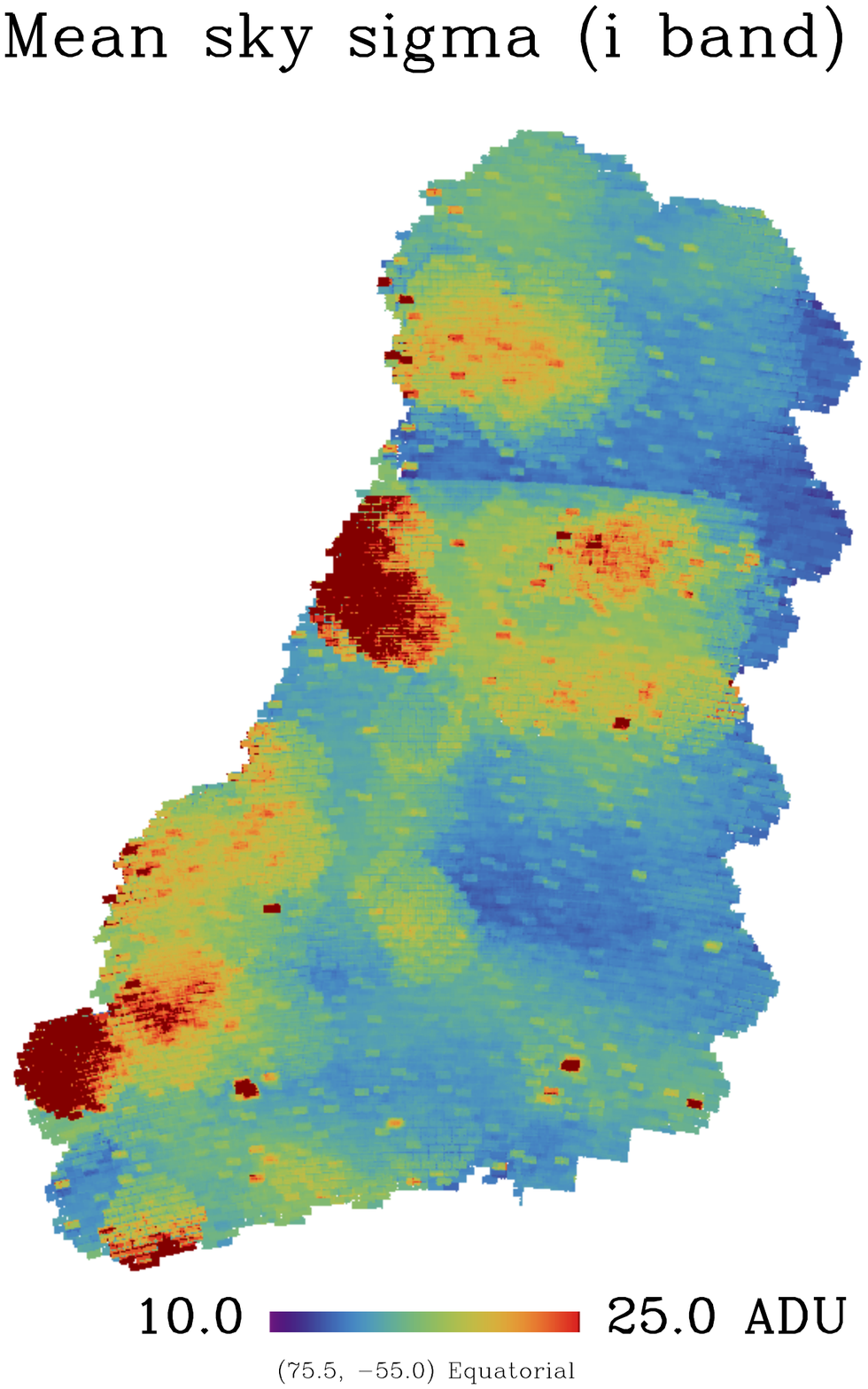}
\\
\includegraphics[width=4cm, trim = 2.9cm 4.0cm 1.3cm 15.9cm, clip]{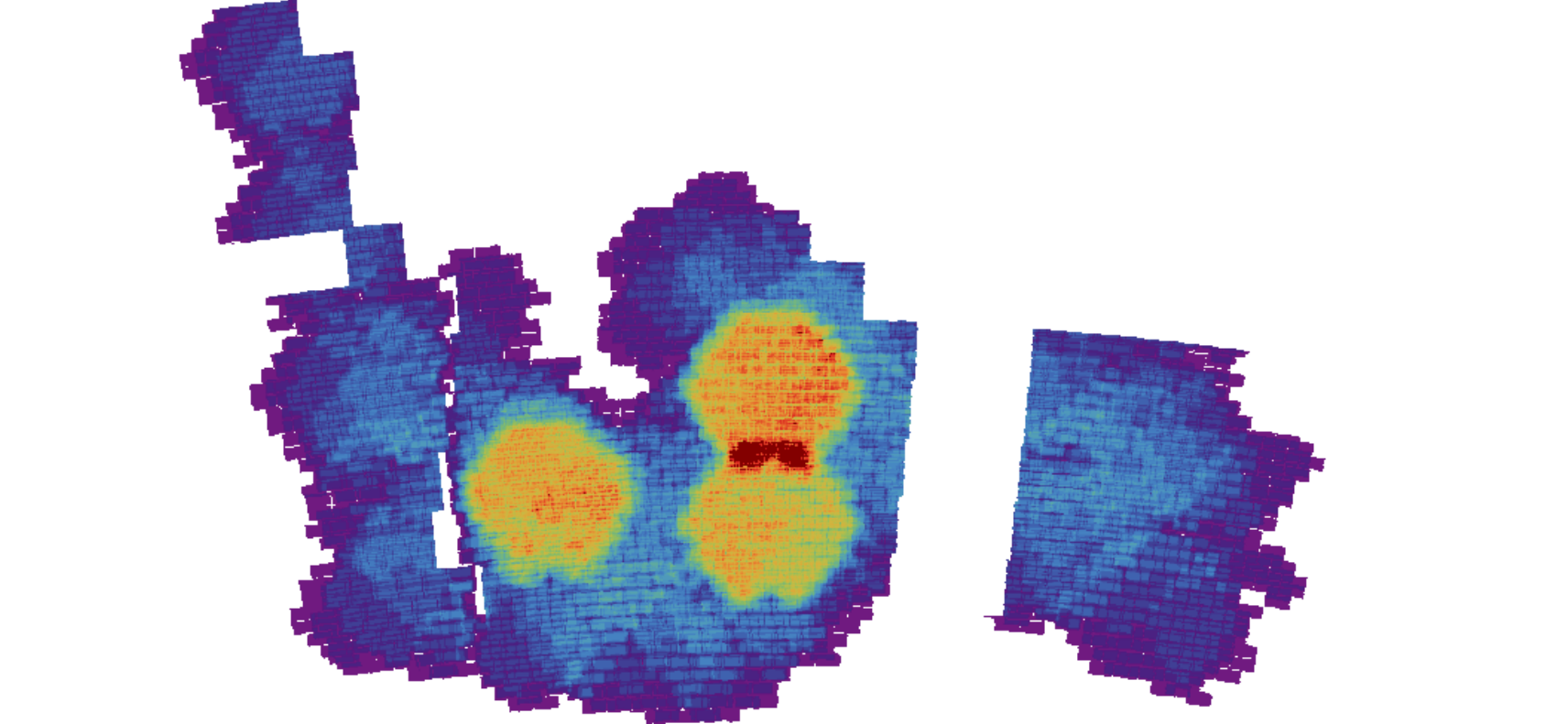}
\includegraphics[width=4cm, trim = 2.9cm 4.0cm 1.3cm 15.9cm, clip]{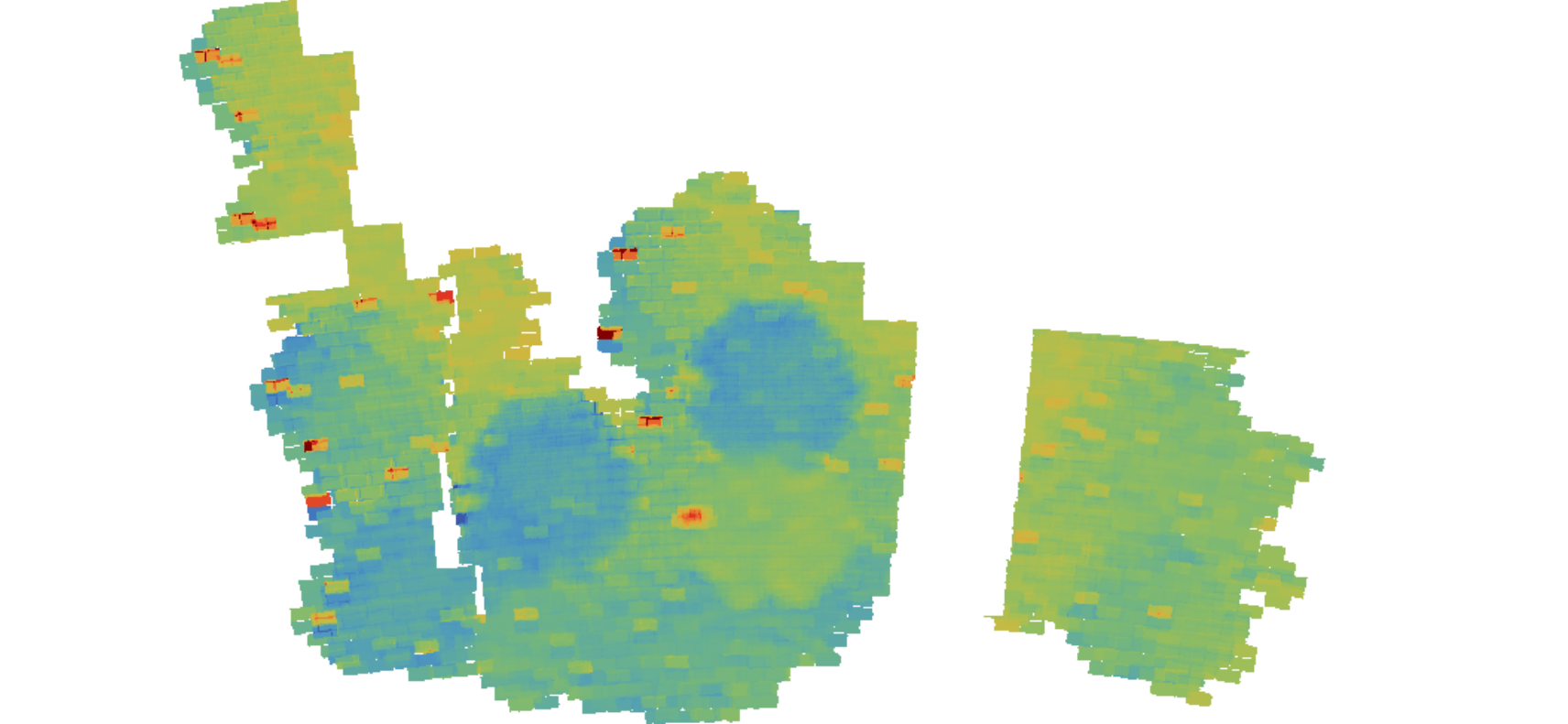}
\\\rule{8cm}{1pt}\vspace{-2mm}
\includegraphics[width=4cm, trim = 2.9cm 3.3cm 1.3cm 0.7cm, clip]{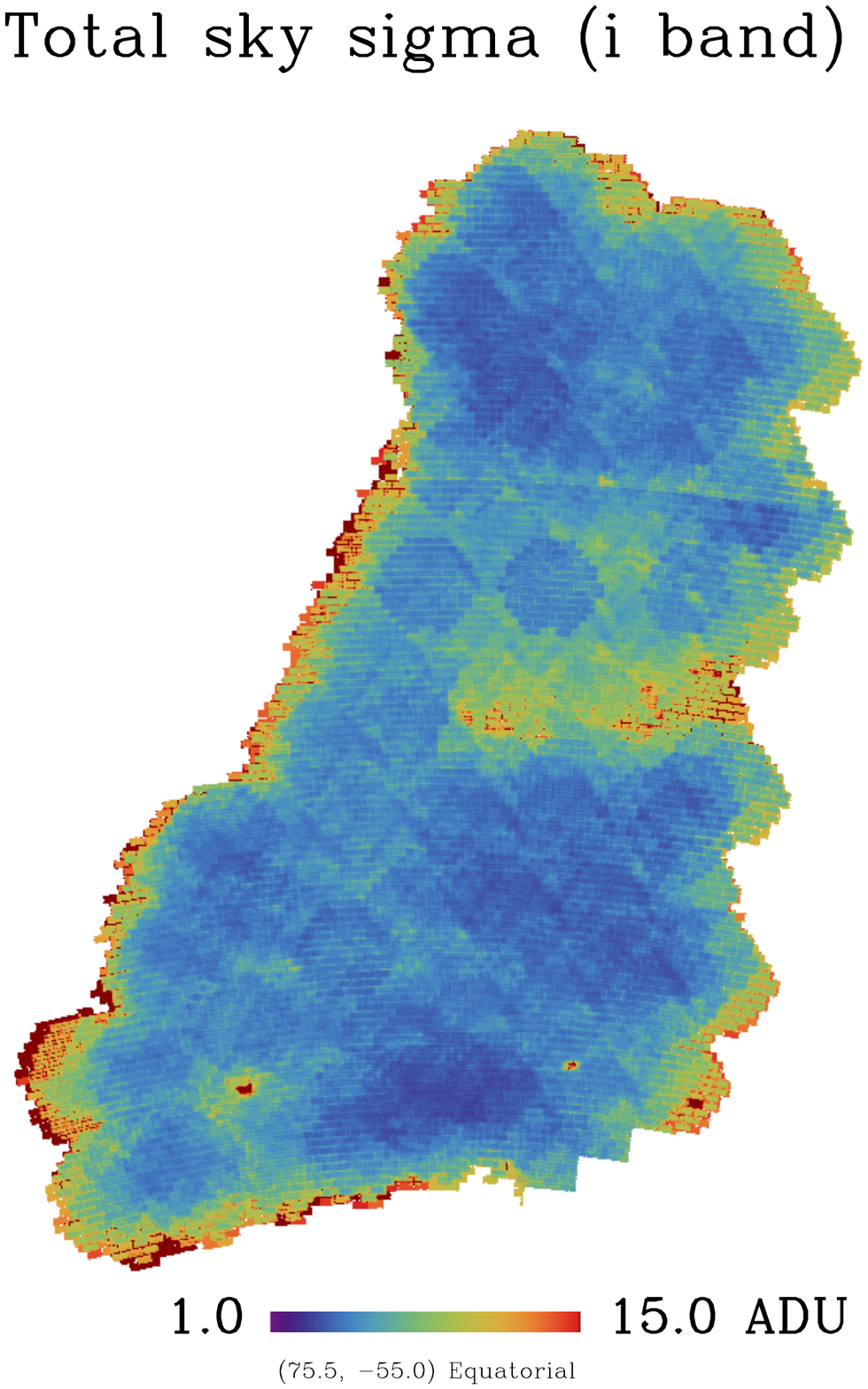}
\includegraphics[width=4cm, trim = 2.9cm 3.3cm 1.3cm 0.7cm, clip]{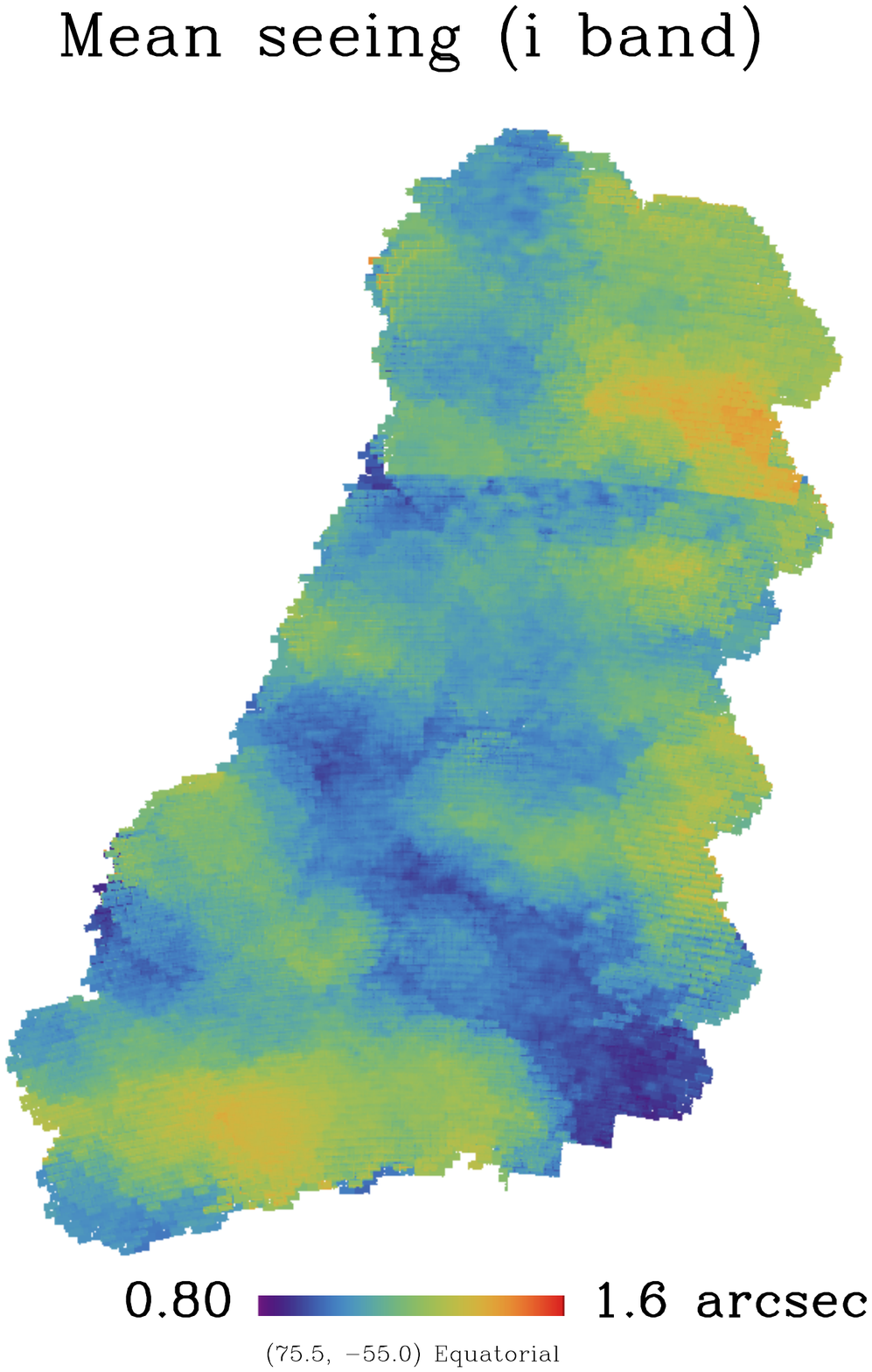}
\\
\includegraphics[width=4cm, trim = 2.9cm 4.0cm 1.3cm 15.9cm, clip]{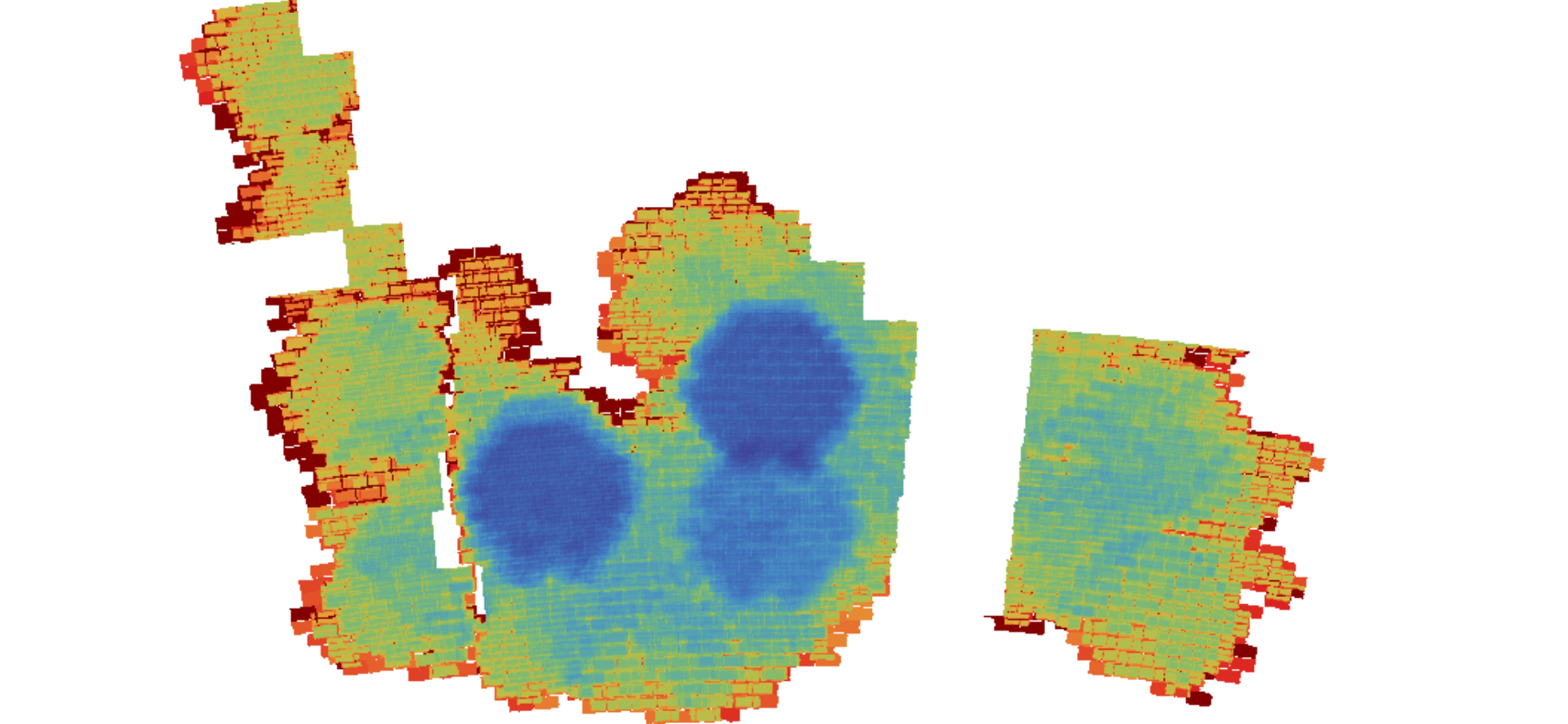}
\includegraphics[width=4cm, trim = 2.9cm 4.0cm 1.3cm 15.9cm, clip]{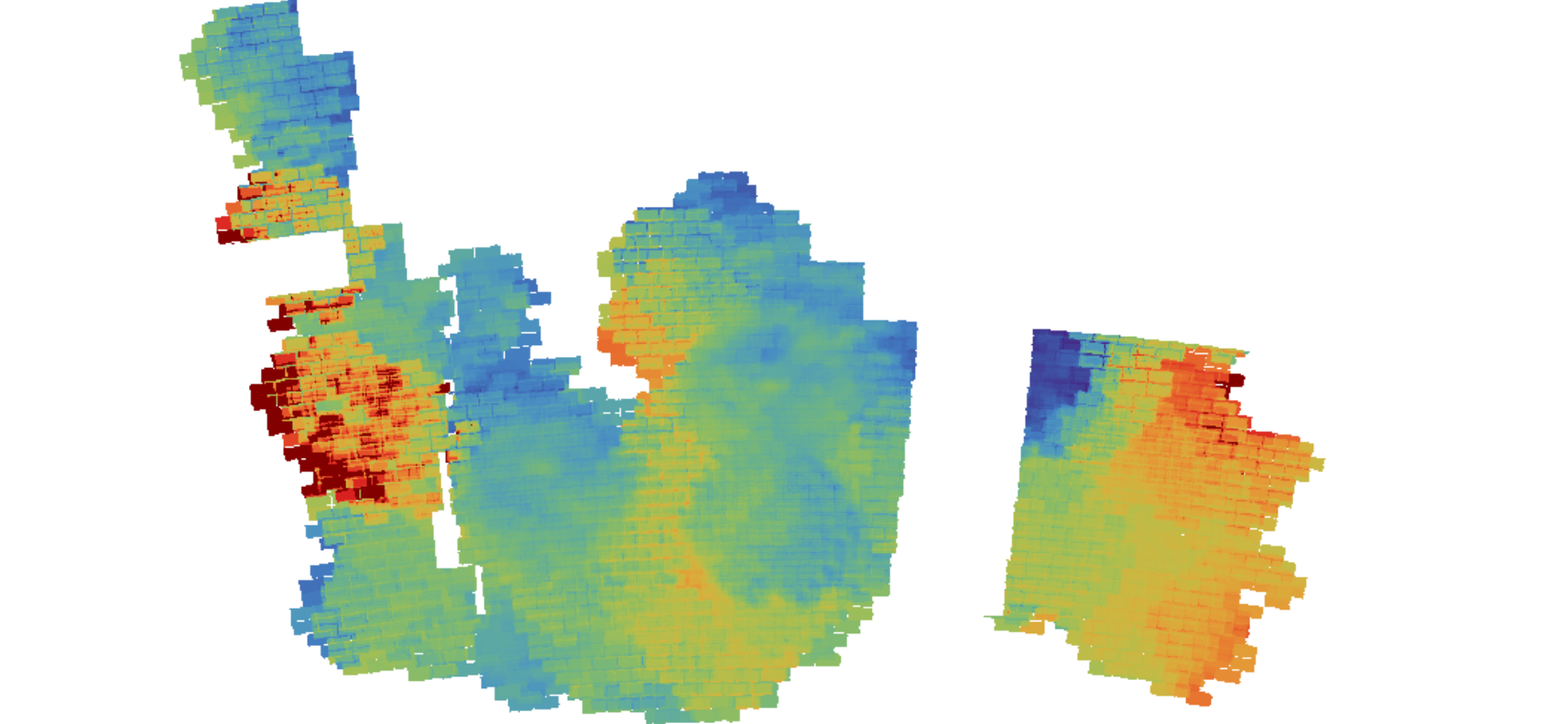}
\\\rule{8cm}{1pt}\vspace{-2mm}
\includegraphics[width=4cm, trim = 2.9cm 3.3cm 1.3cm 0.7cm, clip]{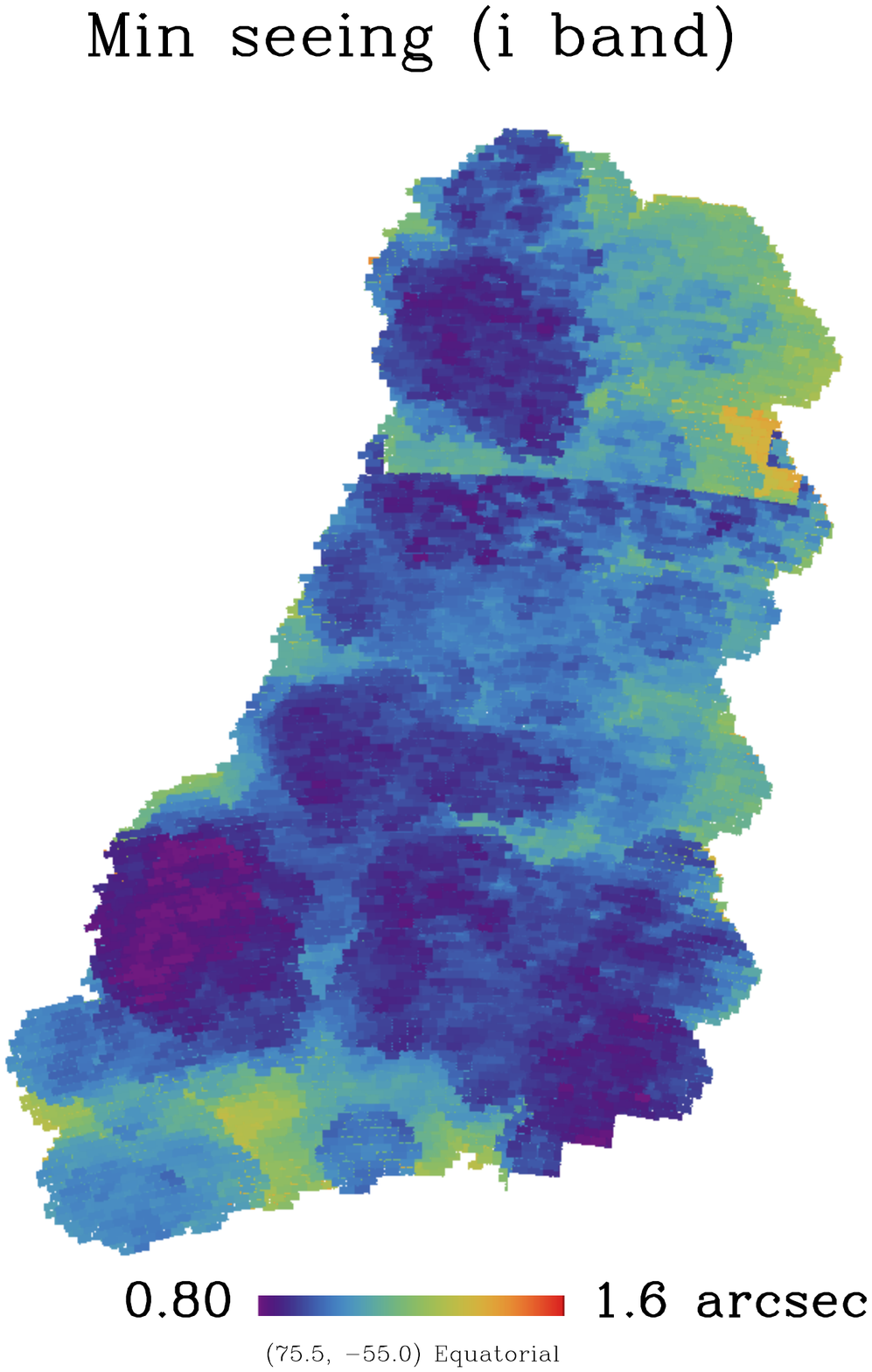}
\includegraphics[width=4cm, trim = 2.9cm 3.3cm 1.3cm 0.7cm, clip]{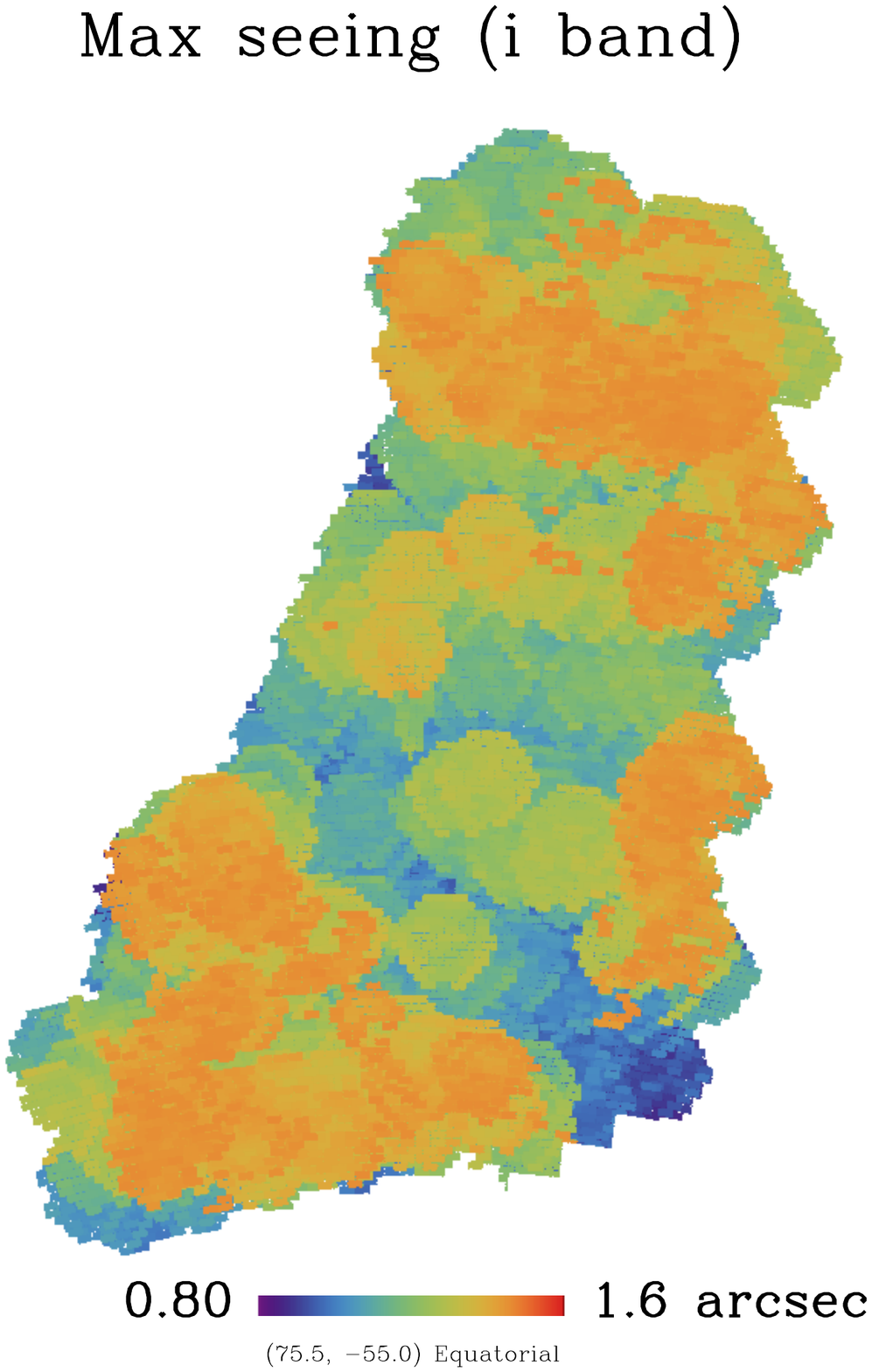}
\\
\includegraphics[width=4cm, trim = 2.9cm 4.0cm 1.3cm 15.9cm, clip]{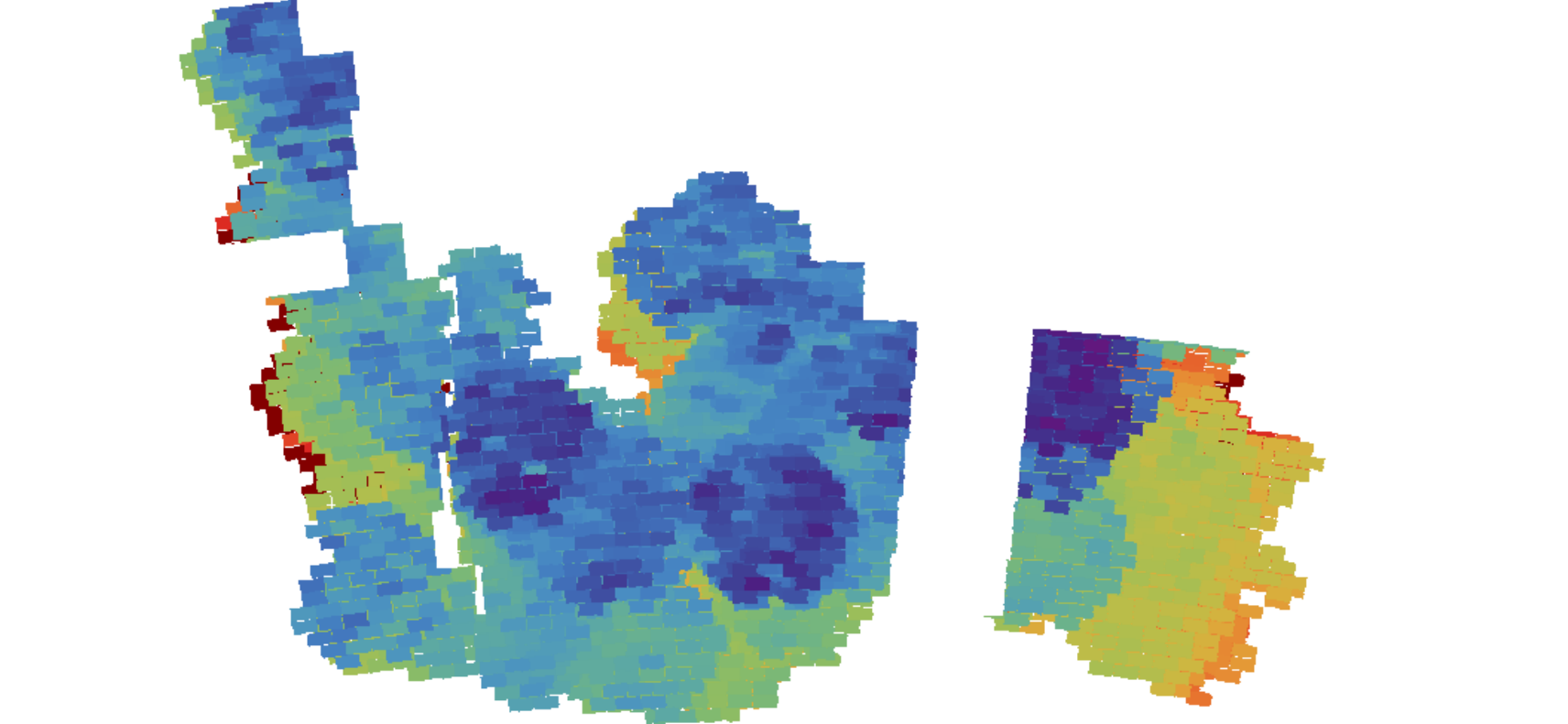}
\includegraphics[width=4cm, trim = 2.9cm 4.0cm 1.3cm 15.9cm, clip]{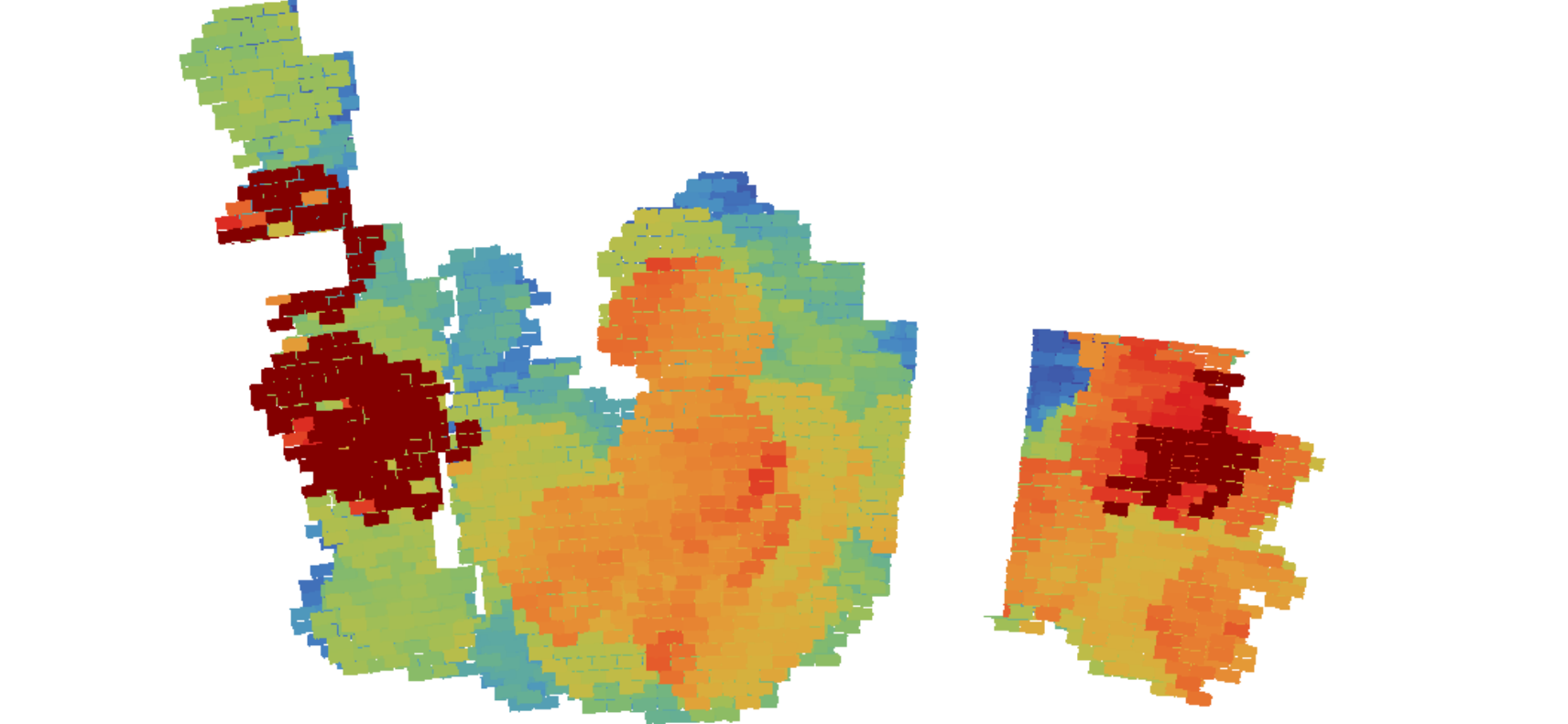}
\caption{Maps of some of the main observational quantities (potential sources of systematics) in the SPT-E and W fields (top and bottom of each sub-panel). \bl{The \healpix maps are produced at $\nside=4096$, where each pixel is the mean value of the observed $\nside=16384$ sub-pixels, in order to obtain more accurate values near the edges of the survey.}  }
\label{fig:sva1maps}
\end{figure}

To create maps for the full SV data set, we create a symbolic tree of CCD and coadd images, resolve their geometries, and project them into \healpix maps. However, unlike the illustration shown in \figref{fig:ccdplot} and \figref{fig:projccdplot}, we now crop the projection to each tile (the black contour of \figref{fig:ccdplot}) since the DESDM software separately processes {tiles using the stacks of CCD images}. We perform this projection in the full SV area and stitch the projected coadds to assemble the full SV footprint. In terms of outputs, we project the following quantities in the five {\it grizY} bands: airmass, seeing, sky brightness, sky sigma (defined later in this paragraph), and exposure time. \bl{These can all affect the quality of the photometric measurements \citep[see \eg][]{TingLi2015descalib}.} We compress the multi-epoch information into average and total maps (\eg mean seeing and total exposure time). For the former, a natural choice would be to take the uniformly-weighted average in each \healpix pixel. However, this choice is probably too simplistic, as in practice images are coadded using weights derived from the flux variance. More precisely, the DESDM pipeline provides a `weight' or `variance' map {for each single-epoch image}. An additional quantity, coined `sky sigma', characterises the variance of the flux in each pixel. For an image $i$ and a given pixel, it is denoted by $\sigma_i$, and depends on a number of parameters, including the flux itself, the gain of the amplifier, the readout noise, the bias correction, and the flat-fielding. Single-epoch images are coadded using these variance maps such that the coadded flux is the weighted average over all exposures,
\equ{
	F_{\rm tot} = \frac{\sum_i w_i p_i F_i }{\sum_i w_i},
}
{in each coadd pixel}, where $w_i = ({p_i^2\sigma_i^2})^{-1}$. The extra $p_i$s are rescaling factors to enforce a common photometric calibration to the single-epoch fluxes. They read
\equ{
	p_i = 100^{ (m_{\rm Z} - m_{\rm Zi})/{5} },
}
where $m_{\rm Zi}$ is the zero point magnitude of the single-epochs and $m_{\rm Z}$ that of the coadd image. The variance of the total flux \bl{in each pixel of the coadd image} is given by
\equ{
	\sigma^2_{\rm tot} = \left[ \sum_i w_i \right]^{-1}.
}
A detailed discussion of these quantities is beyond the scope of this paper, but we note that the total sky sigma is proportional to the magnitude limit of the survey. In the above formulae, we omitted the pixel indexing in $\sigma_i$, but the coadding and the evaluation of $\sigma^2_{\rm tot}$ must be performed pixel by pixel across the coadd image. The technicalities of this process (including the projection and coadding) are handled by the SWARP software \citep{2002ASPC..281..228B}. Yet, the projection formalism presented above can be used to quickly estimate $\sigma^2_{\rm tot}$ (and for example construct approximate magnitude limit maps). For this purpose we compute an average sky sigma per single-epoch CCD image, defined as the pixel average of $\sigma_i$ across the CCD. Rather than computing $\sigma_i$ and $\sigma^2_{\rm tot}$ per pixel, we only need to calculate $\sigma_i$ per CCD  and $\sigma^2_{\rm tot}$ in the distinct regions of image overlap, as shown in \figref{fig:ccdplot}. This yields a significant reduction of the complexity of the full projection, which needs to be performed for the five bands for a number of quantities of interest, using several hundred thousands of single-epoch images. Finally, any quantity of interest can be averaged using the same weights $w_i$ (which we call `sky sigma weights'), which is more useful than the unweighted average. The effective seeing of the coadd images is better approximated by the sky sigma-weighted mean since the coadds are based on these weights. 

A number of maps were constructed for the DES-SV data, in order to capture the spatial fluctuations of the observing conditions and other observational quantities. They are used in numerous SV analyses to perform spatial null tests with the data \citep[\eg][]{2015arXiv150403002V, Crocce2015dessvclustering, Jarvis2015svshear, Becker2015svshear, Giannantonio2015cmbxlssdessv}. \figref{fig:sva1maps} shows some of the main maps for the $i$ band: the total exposure time, the mean sky sigma and total sky sigma, and the minimum, maximum and mean seeing. All quantities were calculated according to the previous scheme, {\ie the weighted average method and the sky sigma weights, with the exception of the mean sky sigma maps. This is because the weighted sky sigma is equivalent to the total sky sigma described above. Showing both maps sheds light on the difference between adding the noise properties linearly or in quadrature. In the following we analyse these maps and detail the implications for the analyses of SV data. We focus on the SPT-E and W regions since they are the largest contiguous regions of the data. }

\subsection{Analysis of the DES SV observing conditions}

\begin{figure}
\includegraphics[width=8.5cm, trim = 0.4cm 0.4cm 0.1cm 0.4cm, clip]{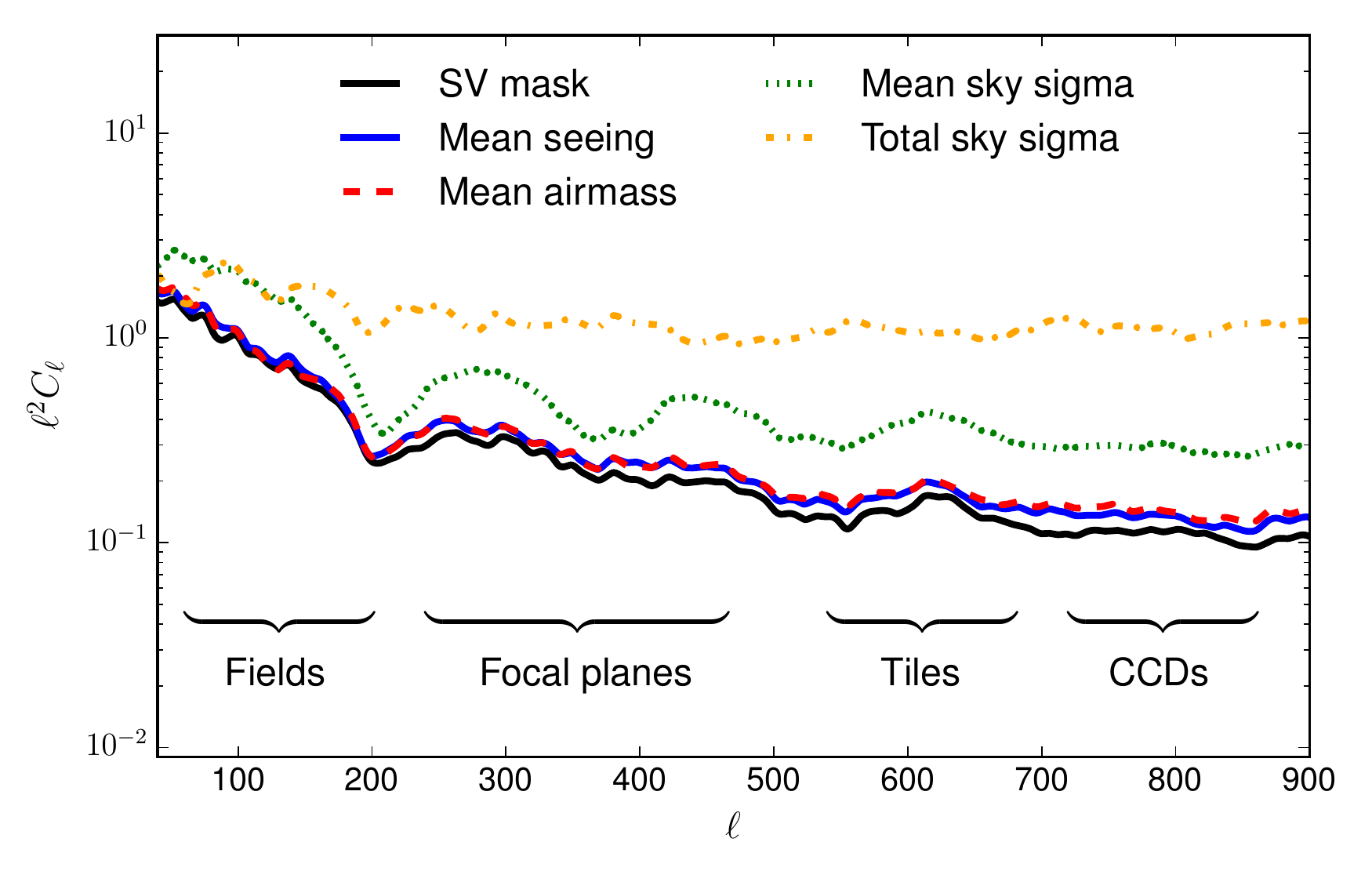}
\caption{Full-sky angular power spectra of some of the $i$ band observational systematics shown in \figref{fig:sva1maps}. Prior to power spectrum estimation, the maps were divided by their average values in order to obtain dimensionless $C_\ell$'s. The power spectrum of the DES-SV coverage mask \bl{(presented in \citealt{Rykoff2015sva1gold, Crocce2015dessvclustering})} is also shown in black. Any excess of power relative to the mask implies structure and features in the maps, which can yield non-trivial contamination and systematics in the galaxy catalogues. We also indicate the characteristic scales affected by the geometry of the SV survey and DECam instrument.}
\label{fig:sys_cls}
\end{figure}

The maps shown in \figref{fig:sva1maps} exhibit significant structure and features on all scales, mostly because DES data have three intrinsic scales on which their properties can vary: {the size of the DECam focal plane ($2.2$ deg diameter field of view), the coadd tile ($0.75\times0.75$ \sqdeg), and the single CCD ($0.3 \times 0.15$ \sqdeg)}. In spite of the random \bl{offsets} and overlap of the focal plane when obtaining images and coadding them, these three scales get imprinted in the projected observing conditions. For example, the focal plane geometry is clearly visible in the total sky sigma maps in a number of regions. This is due to a significantly lower or greater number of observations, or to their respective noise levels (sky sigma). Also, the mean seeing map is affected by outliers, \ie by extreme (low or high) values of seeing in the set of single-epochs, as shown in the min/max maps in the bottom of \figref{fig:sva1maps}. The rectangular CCD geometry is also visible in the maps, especially near the edges. In addition, the observing properties of the 62 CCDs in a given single-epoch are very correlated since they experience quasi-identical observing conditions. By contrast, correlations between exposures are due to proximity in time, for example if the observations were taken the same night. Finally, the tiles edges are particularly visible in truncated regions or due to applying different zero point magnitudes (\eg the centre of SPT-W, or the sharp transition in the upper part of SPT-E).

To identify which scales may be affected by the features described above, we compute the full sky angular power spectra of the maps in \figref{fig:sva1maps} (the full SV, not only the SPT-E and W regions). The results are shown in \figref{fig:sys_cls}; all spectra are made dimensionless and normalised such that $\sum_\ell C_\ell=1$ to clarify the comparison. As seen before, all maps exhibit significant power on all scales. The labels show which multipole ranges correspond to the typical scales of the SV fields, DECam focal plane, tiles, and CCDs. It is important to note that many of the features of \figref{fig:sys_cls} are due to the sky coverage (\ie the footprint) of SV, not the correlations in the observed regions. This is emphasised by an extra line showing the power spectrum of the DES-SV footprint mask. \bl{Here we do not deconvolve the effect of the mask on the power spectra because it typically redistributes the power between the $\ell$ modes. In the pseudo-spectrum estimation method, this deconvolution assumes flat priors on the power spectra, while quadratic maximum likelihood estimators can incorporate more flexible priors on the power spectra \cite[see \eg][]{Leistedt2013excessdr6}. This deconvolution would significantly affect the observed power spectra due to the small sky coverage of SV data. By contrast, not deconvolving the mask enables one to separate the scales affected by the survey coverage and by the observing conditions.} The significant power in the $\ell\in[0,200]$ range is mostly due to the size and shape of the SV fields (all fields except SPT-E and W have approximately the size of the focal plane). In the other power spectra, any power in excess of the black line is due to structure within the fields, \ie to the features described previously. As expected, airmass and seeing maps mostly have additional power on small scales. But the sky sigma maps have much more power on all scales, in particular around the focal plane and coadd scales. 

\begin{figure}\hspace*{-2mm}
\includegraphics[width=8.9cm, trim = 0.5cm 2.0cm 0.3cm 1.0cm, clip]{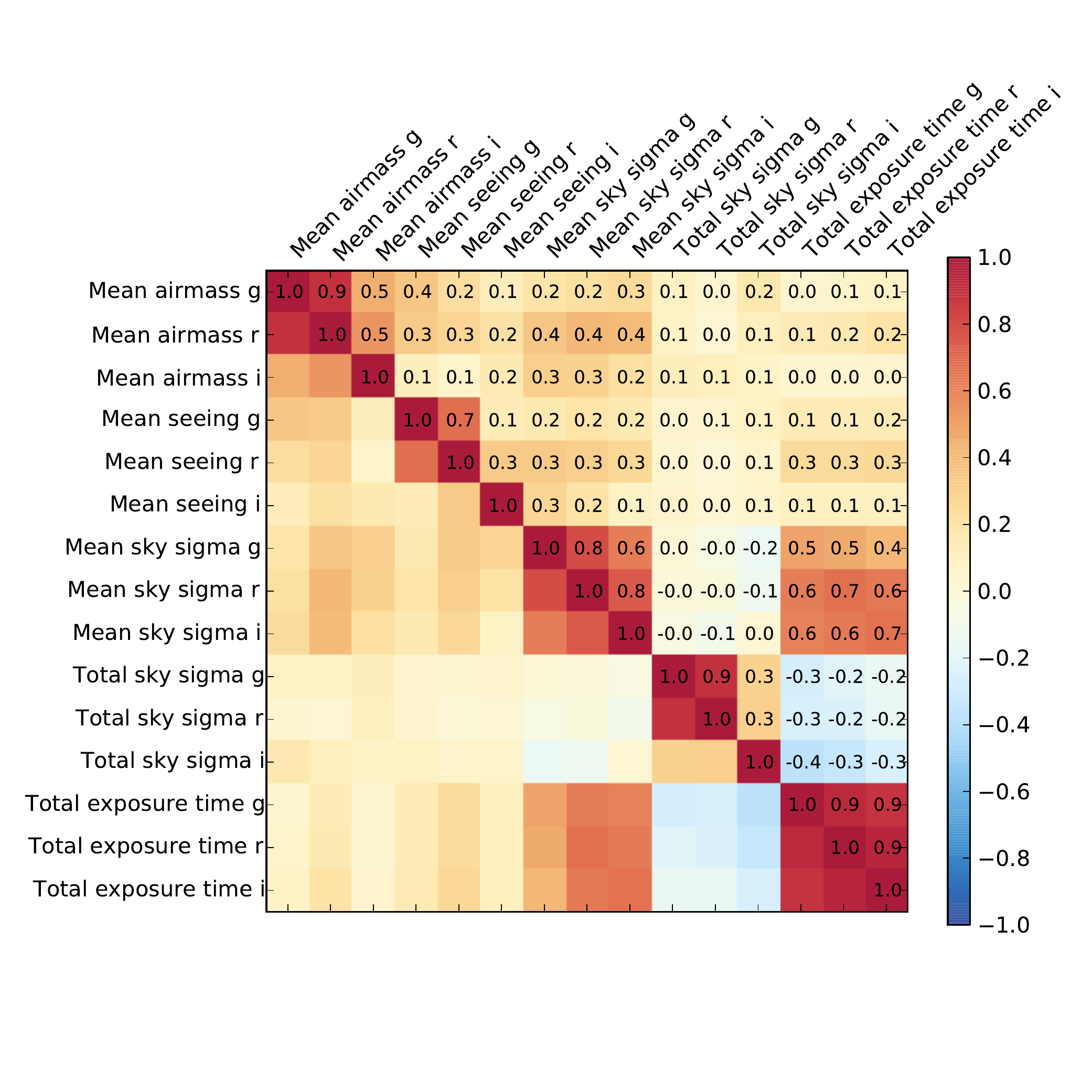}
\caption{Correlation coefficients between some of the maps produced for the DES-SV data.}
\label{fig:syscorr_sva1}
\end{figure}

As seen in \figref{fig:sva1maps}, the maps of the observing conditions are correlated. \figref{fig:syscorr_sva1} shows the Pearson correlation coefficients of the DES-SV maps in the $gri$ bands (calculated for the full SV area). These spatial correlations have two origins: the time correlations between observations made closely spaced in time, and physical correlations between some of the properties. For example, the noise level and seeing are correlated. 
 
In conclusion, the observing conditions fluctuate significantly on a wide range of scales, and may affect the properties of the galaxies detected in DES coadd images. Any resulting spurious spatial correlations that propagate into the galaxy catalogues will need to be detected and eliminated. \bl{Typical techniques to mitigate these effects in clustering analyses include modelling the survey window function \citep[\eg][]{1996MNRAS.283.1227M, 2010MNRAS.406..803B}, or using cross-correlations \citep{2002ApJ...579...48S, ross2011weights, ross2012systematics, ho2012cosmoweights, Crocce2015dessvclustering} or mode-projection \citep{Leistedt2013excessdr6, Leistedt:2014wia} to correct or mask the spatial modes affected by the observing conditions}. Crucially, these approaches require the availability of accurate templates of the sources of systematics, which were precisely constructed in this section. We now turn to a concrete example of use of these templates.

\section{Application to BCC-UFig}

\begin{figure}\centering
\includegraphics[width=2.7cm, trim = 3.9cm 3.3cm 2.7cm 3.9cm, clip]{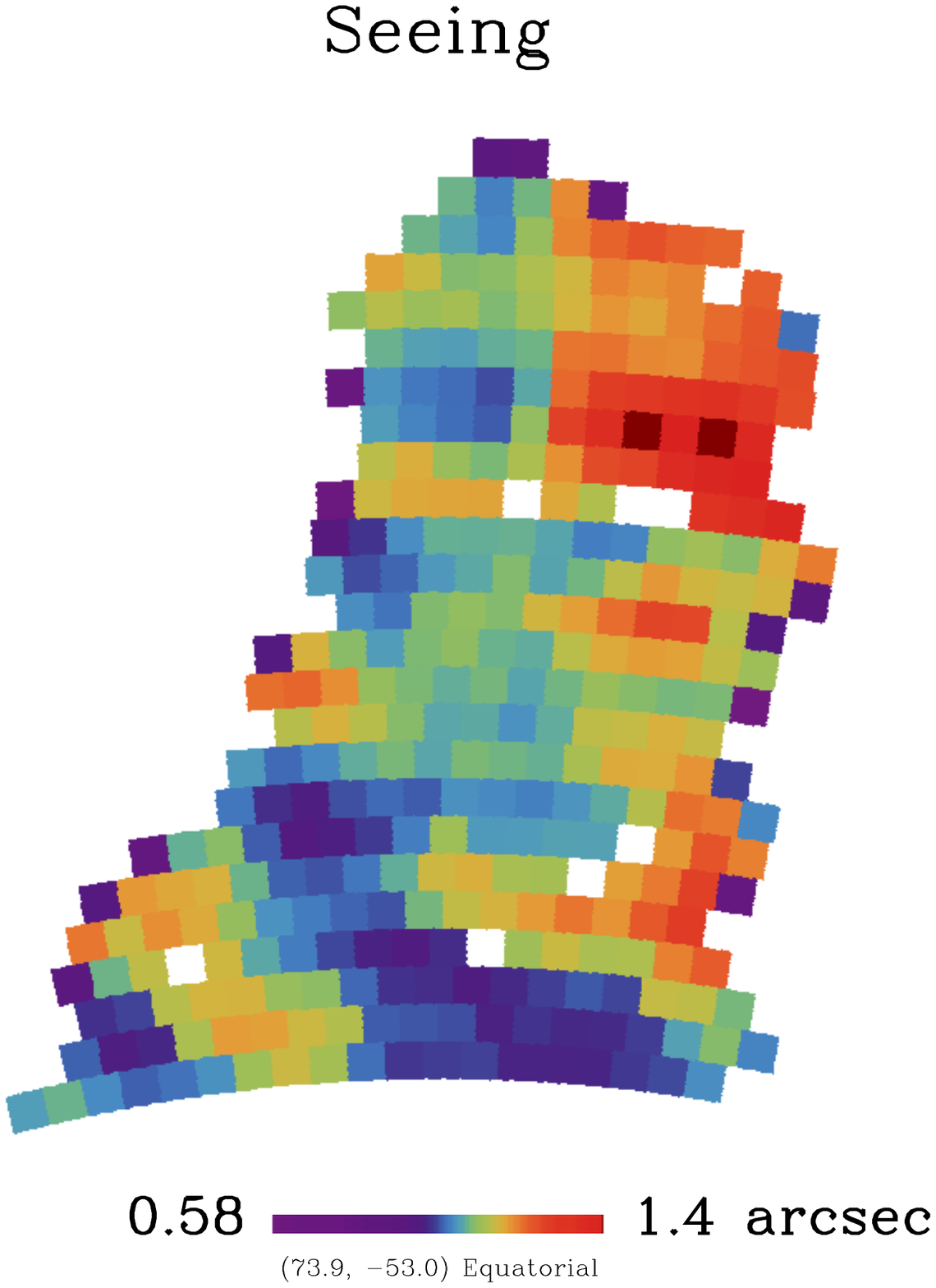}
\includegraphics[width=2.7cm, trim = 3.9cm 3.3cm 2.7cm 3.9cm, clip]{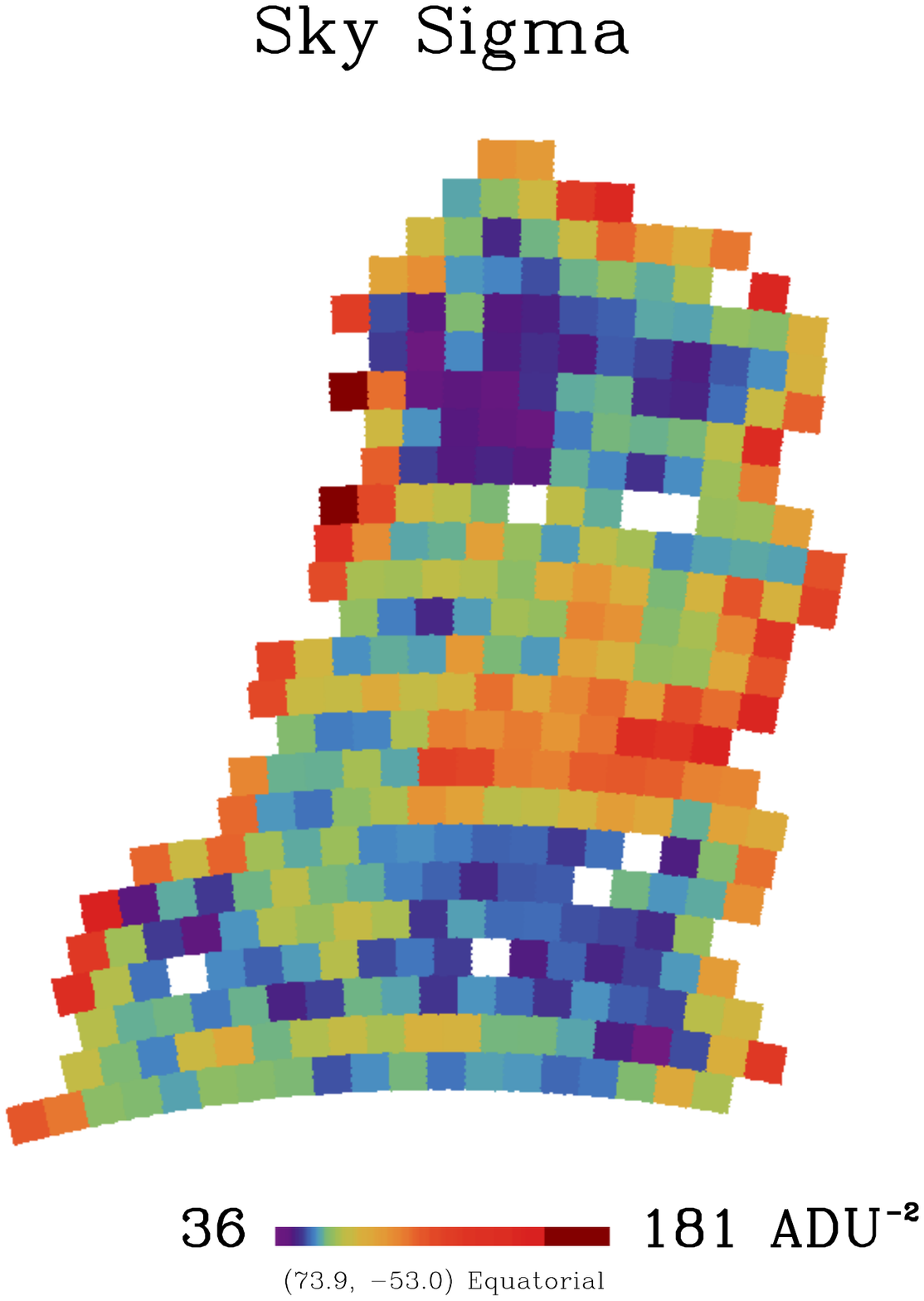}
\includegraphics[width=2.7cm, trim = 3.9cm 3.3cm 2.7cm 3.9cm, clip]{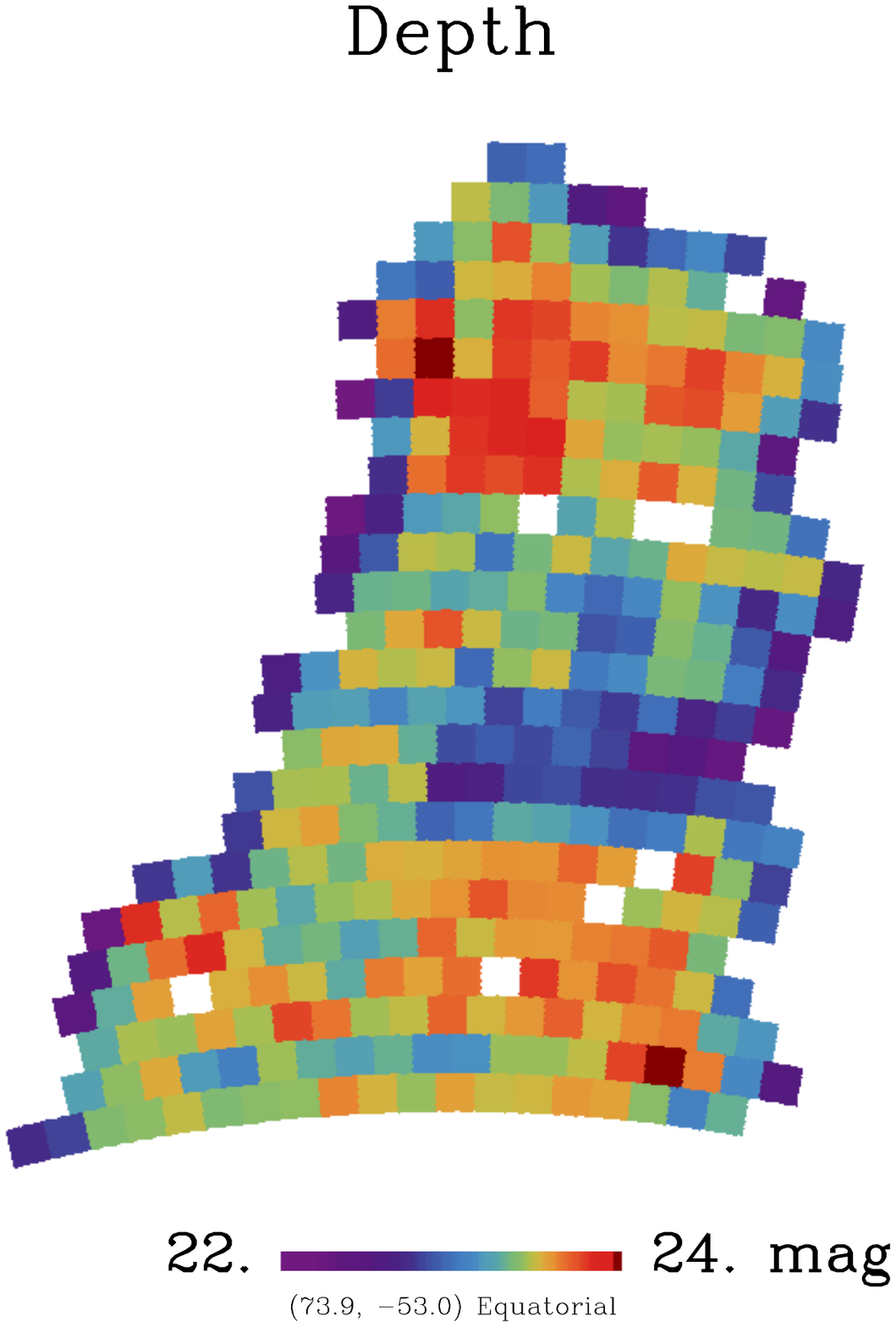}
\includegraphics[width=2.7cm, trim = 3.9cm 3.3cm 2.7cm 3.9cm, clip]{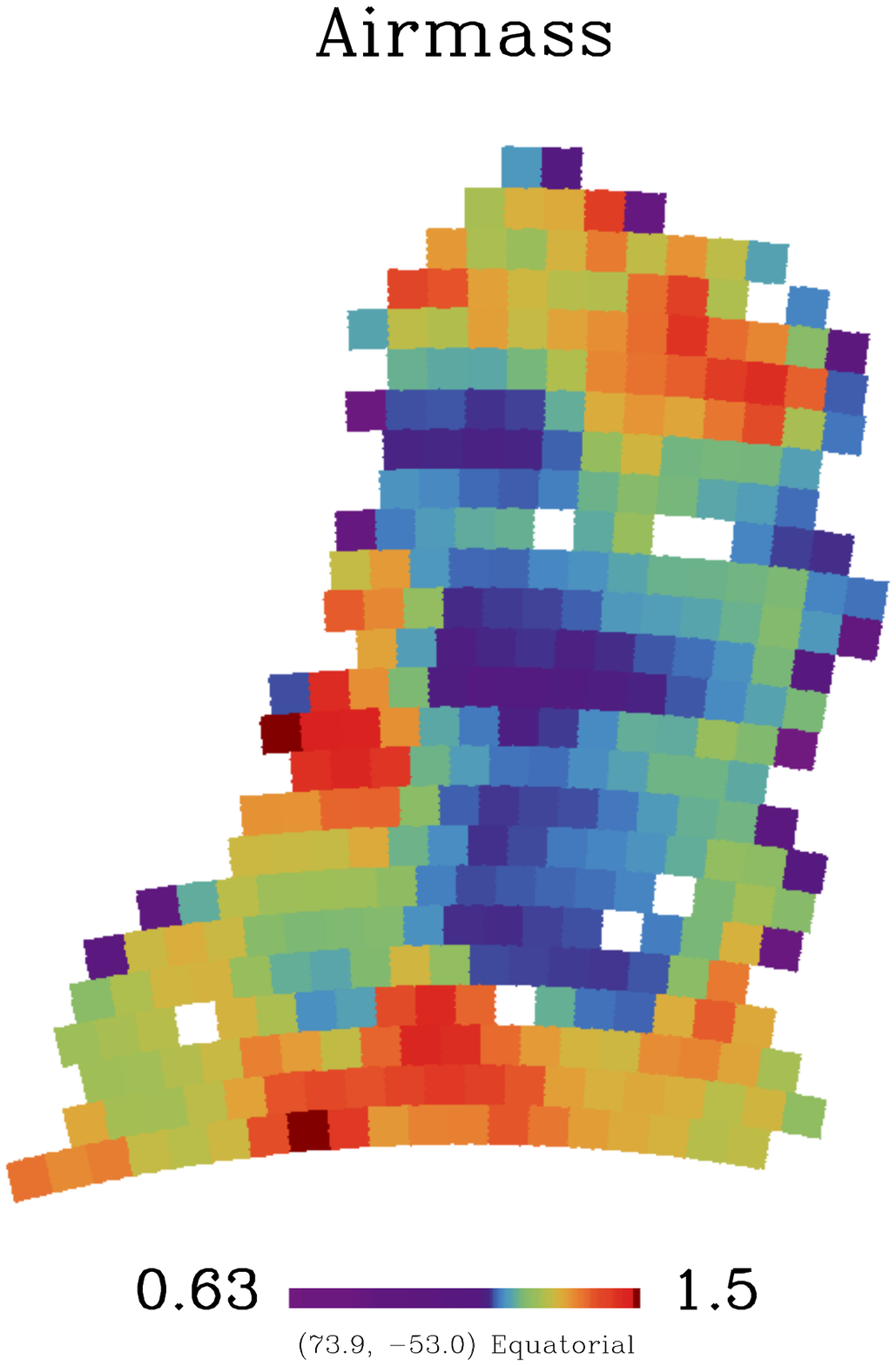}
\includegraphics[width=2.7cm, trim = 3.9cm 3.3cm 2.7cm 3.9cm, clip]{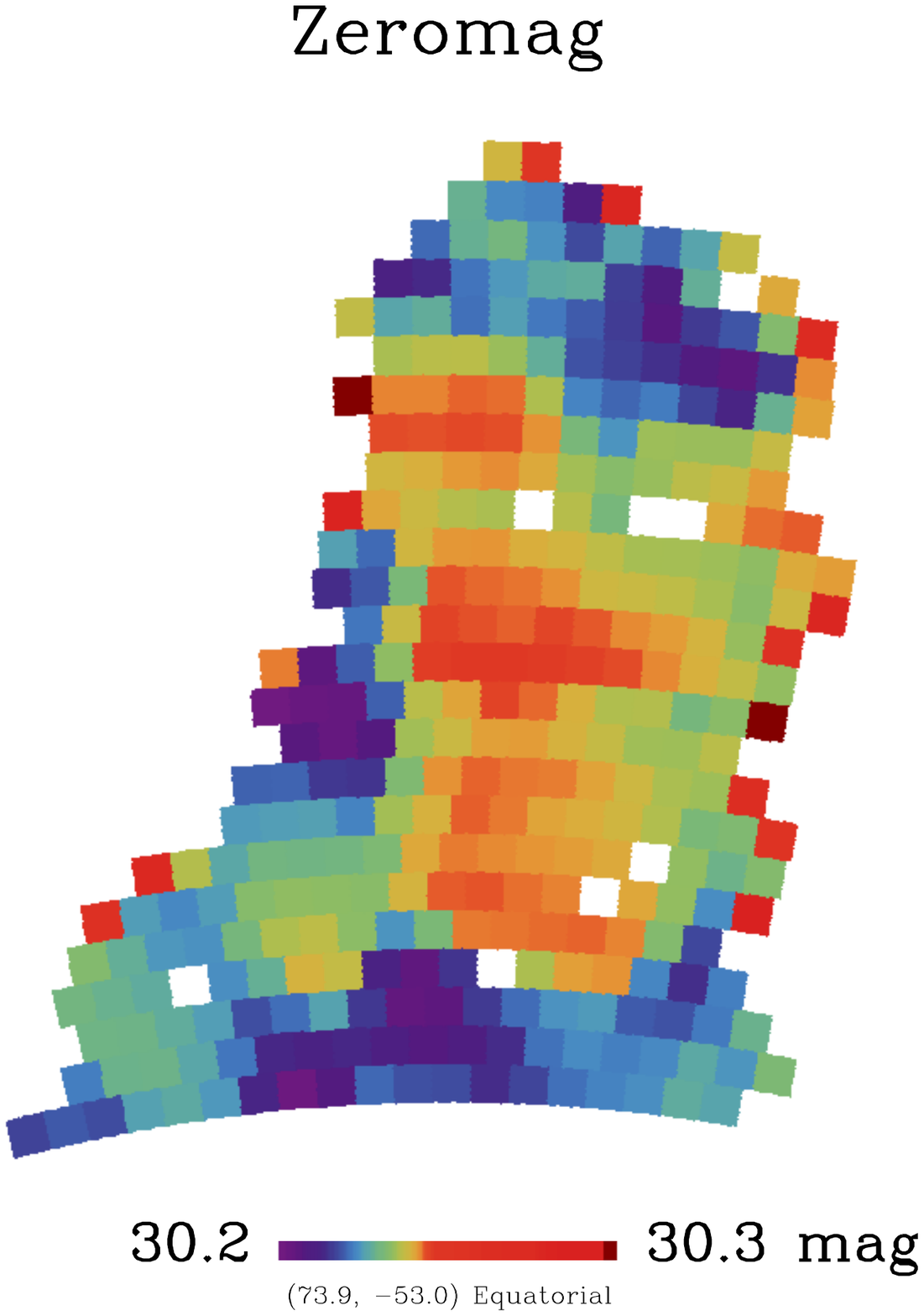}
\includegraphics[width=2.7cm, trim = 3.9cm 3.3cm 2.7cm 3.9cm, clip]{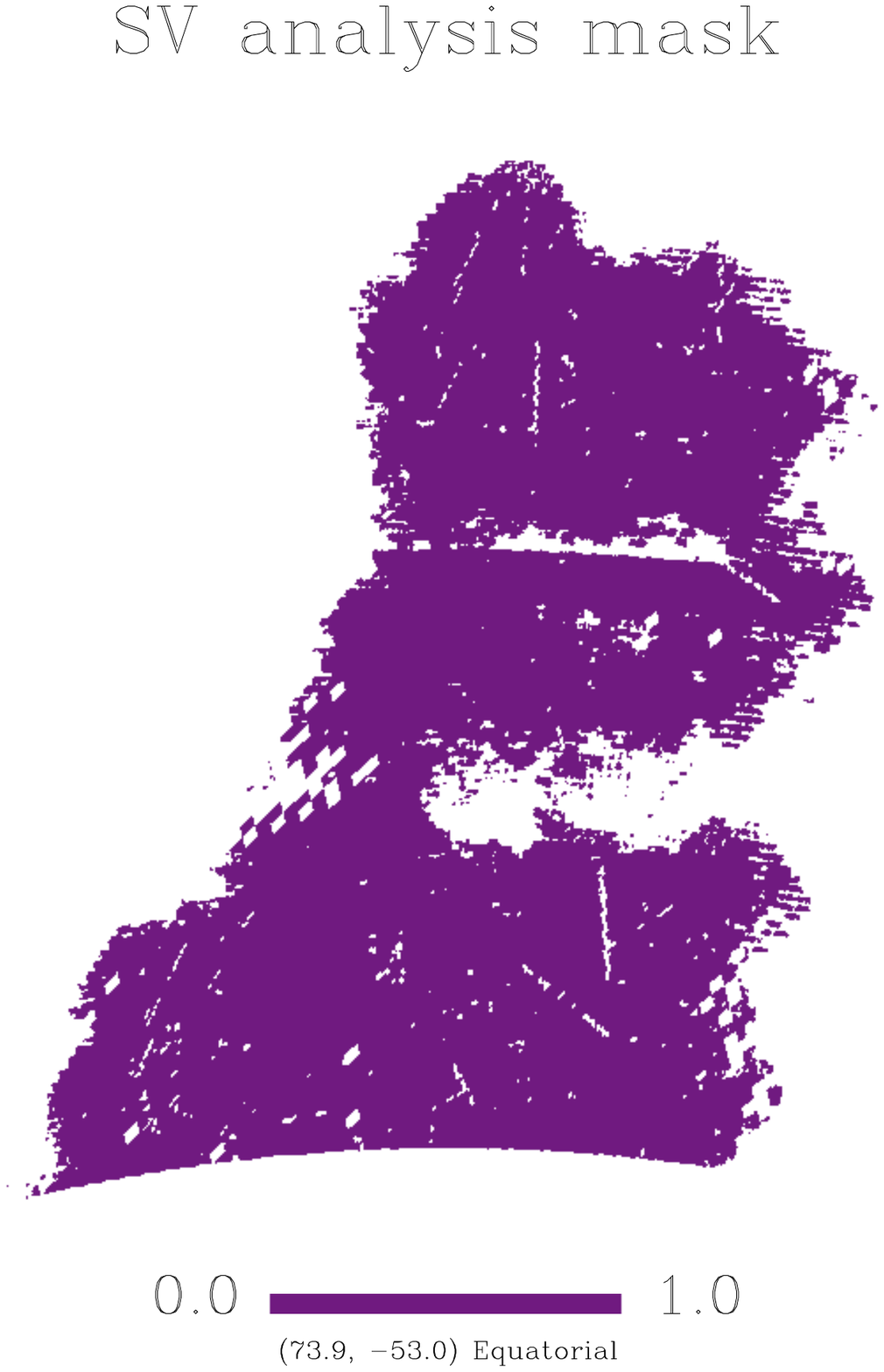}
\caption{Maps of the $i$ band SV observing conditions incorporated in the BCC-UFig simulation, obtained by smoothing the maps of \figref{fig:sva1maps} in tiles. The analysis mask shows the region considered when measuring the redshift distributions and galaxy number densities presented below.}
\label{fig:ufigmaps}
\end{figure}

The BCC-UFig \citep{Chang2014bccufig} is a framework of image-level simulations of the DES-SV data. It relies on the Ultra Fast Image Generator (UFig, \citealt{Berge2013ufig}) and the BCC cosmological simulations (the Blind Cosmology Challenge, \citealt{busha2013bcc}) in order to obtain realistic images of a galaxy survey simulated with a known cosmological model. The BCC-UFig covers the SPT DES-SV region, and consists of 480 coadd images in the {\it griz}-bands, and 432 in the {\it Y}-band. As detailed in \cite{Chang2014bccufig}, these images were processed using the same software packages as the DESDM SVA1 pipeline. In this paper we exploit the fact that the simulated BCC-UFig images integrate some of the actual observing conditions of the DES-SV data. In particular, the simulated coadd images incorporate the median values of the seeing, limiting magnitude, and magnitude zeropoint of the true DES-SV images. These quantities were obtained from the products presented in the previous section ({\ie maps of the median observing conditions}, analogous to the mean maps shown in \figref{fig:sva1maps}), by averaging each map over the surface of the tiles. The result of this smoothing is shown in \figref{fig:ufigmaps}, and we comment on its effect on the galaxy catalogues below.  \bl{The fact that the BCC-UFig is based on simulated coadd images and not on single-epochs is the main difference with the real SV data. However, as discussed below, BCC-UFig reproduces most of the spatial systematics found in the data and relevant to clustering and weak lensing analyses, because these are due to fluctuations in observing conditions at scales larger than the coadds.}

Survey simulations like the BCC-UFig can be used to test analysis techniques and pipelines in the presence of realistic systematics. For example, \cite{Chang2014bccufig} used the BCC-UFig to compare the performances of various star-galaxy classifiers and study the evolution of the observed galaxy and stellar densities as a function of some of the observing conditions (depth, seeing, Galactic latitude). Such tests cannot be performed at high significance in the real data due to the small size and sky coverage of the sample of spectroscopically confirmed galaxies (based on the COSMOS and SN fields, shown in red in \figref{fig:sva1_tiles}). 

In this paper,  we produce galaxy catalogues based on the BCC-UFig and compare them with the SV data catalogues. We mostly attempt to mimic the galaxy catalogues used in the clustering and cross-correlation analyses of SV \cite{Crocce2015dessvclustering, Giannantonio2015cmbxlssdessv}. We first construct a multi-band catalogue by cross-matching the positions of the objects detected in the $griz$ bands. We then remove all objects with extreme fluxes or colors: $x>30$ and $x-y>3$ or $x-y<-1$, where $x$ and $y$ are {\tt mag\_auto} magnitudes in $griz$ bands measured by SExtractor. To select galaxies in this catalogue, we use the `modest' classifier. As described in \cite{Chang2014bccufig, Soumagnac2013dessgsep}, objects are labelled as galaxies by this classifier if they do \textbf{not} satisfy any of the following criteria: ({\tt mag\_auto\_}i $< 18$ and {\tt class\_star} $> 0.3$) or ({\tt spreadmodel\_}i + 3*{\tt spreadmodel\_err\_}i$<0.003$) or ({\tt mag\_auto\_}i $< 21$ and {\tt mag\_psf\_}i $> 30$). Finally, we only consider objects with $18<$ {\tt mag\_auto\_}i $< 22.5$ in the SPTE region, and we split this galaxy sample into redshift bins using photometric redshifts. {We now investigate the realism of these galaxy samples, first in terms of their redshift distributions.}

\subsection{Photometric redshifts and redshift distributions}

Photometric redshifts (\photoz) --- redshifts estimated from broad-band fluxes and colours --- are one of the main sources of uncertainties in imaging surveys, and it is essential to reproduce this aspect of the data with BCC-UFig galaxies. We employ three \photoz codes: BPZ \citep{benitez2000BPZ, Coe2006BPZ}, TPZ \citep{Kind2013TPZ, Kind2013TPZ2}, and ANNz2 \citep{Sadeh2015ANNZ2}. These rely on very distinct algorithms that were tested on early SV data in \cite{sanchez2014dessva1photoz}. They are also used in the main SV clustering and cosmic shear analyses \citep{Giannantonio2015cmbxlssdessv, Crocce2015dessvclustering, Jarvis2015svshear, Becker2015svshear}. For details on the three codes and a updated comparison using SV data, we refer the reader to \cite{Bonnett2015photozforWL}. Here we only provide a brief summary of the three algorithms and focus on comparing the redshift distributions inferred from the SV data and the BCC-UFig simulation.
  
\begin{figure}
\hspace{-3mm}\includegraphics[width=8.7cm, trim = 0.0cm 1.3cm 1.5cm 1.5cm, clip]{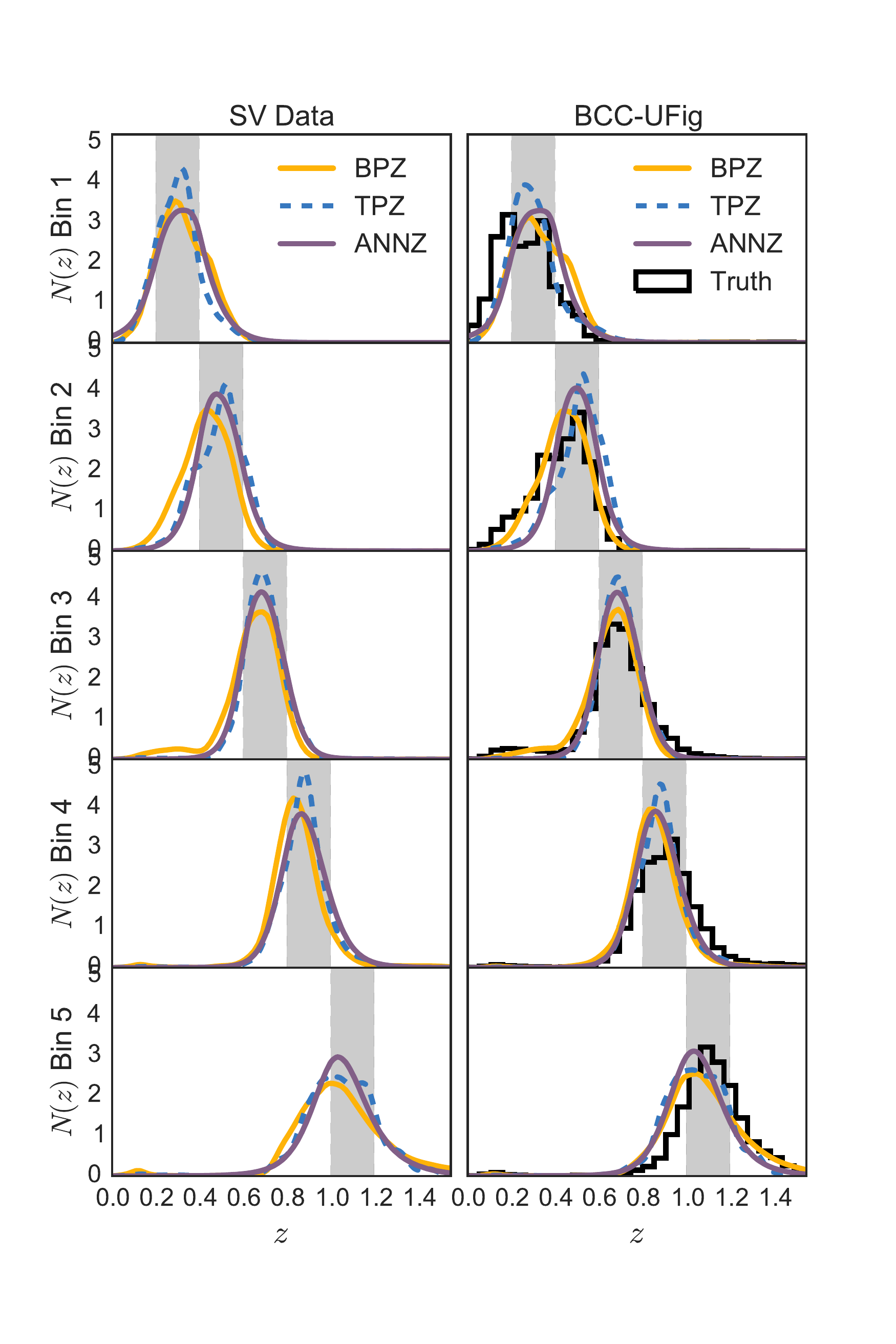}
\caption{Redshift distributions of the SV data (left) and BCC-UFig (right) catalogues obtained using the photometric redshift estimation methods trained on a spectroscopic sample of galaxies (see text for details). \bl{They are normalised such that $\int N(z)dz=1$. By comparison, the variance obtained by randomly splitting the redshift samples is of the order of $0.01$ in all redshift bins.}}
\label{fig:dndz}
\end{figure}

\begin{figure}
\hspace*{-2mm}\includegraphics[width=8.7cm, trim = 0.0cm 1.3cm 1.5cm 1.5cm, clip]{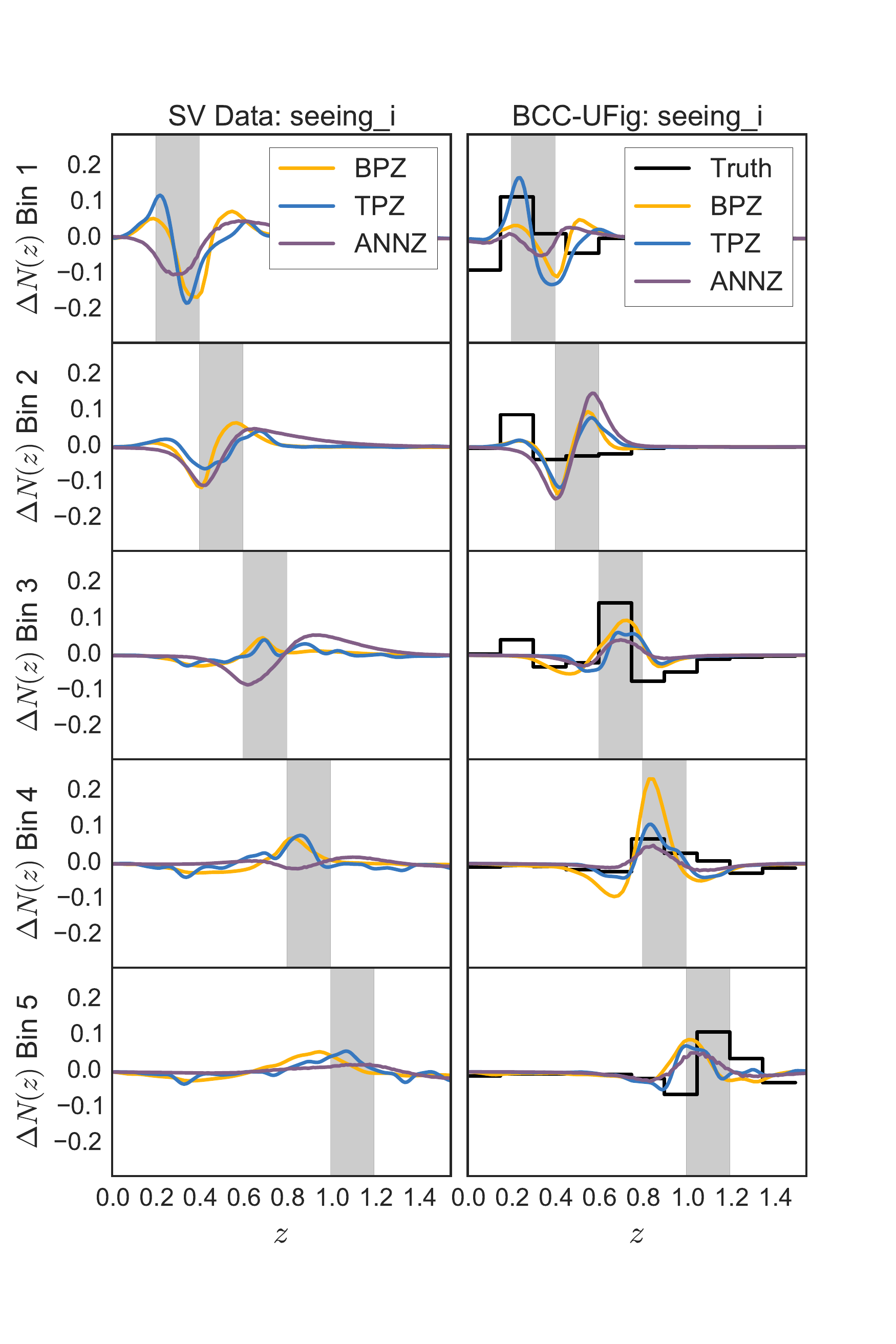}
\caption{Difference between the redshift distributions for {mean seeing} bins, \ie computed for the good (low seeing) and bad (high seeing) regions.}
\label{fig:dndzdiff_seeing}
\end{figure}

BPZ is a Bayesian template fitting \photoz code that relies on a set of calibrated template spectra, which are redshifted and converted into template colours using the DES filters. It computes a posterior probability for the redshift of each object given its observed colours and errors, by fitting for all templates and marginalising over the choice of template. By contrast, TPZ and ANNz2 are machine learning codes that must be trained on a representative sample of the data to infer a set of heuristic rules (\ie a flexible data-driven model) to compute the redshift from the observed photometric colours. TPZ is a publicly available code\footnote{http://lcdm.astro.illinois.edu/code/mlz.html} based on prediction trees and random forests, while ANNz2 uses a combination of machine learning algorithms, including neural networks and $k$-nearest neighbours. The three photo-$z$ codes deliver a redshift probability distribution function (PDF) and a photo-$z$ point estimate, usually measured as the mean or the mode of the PDF.

We employ the BPZ, TPZ and ANNz2 algorithms that were trained and calibrated on the SV data, more specifically on the sample of galaxies presented in \cite{Bonnett2015photozforWL} {for which spectroscopic redshifts are also available (about $46,000$ galaxies)}. This sample is shown in red in \figref{fig:sva1_tiles} and was used to calibrate the BPZ template prior and train the TPZ and ANNz2 methods. Note that we only use {\tt mag\_auto} magnitudes and colours with BPZ and ANNz2, and {we only include the magnitude errors in the training of TPZ (where they are used to perturb the magnitudes when re-training the prediction trees, in order to obtain reliable redshift posterior PDFs)}.

Following most analyses of SV \citep[\eg][]{Crocce2015dessvclustering, Giannantonio2015cmbxlssdessv}, we create five BCC-UFig redshift samples by selecting the objects with photometric redshift falling in a top hat window of size $\Delta z = 0.2$ in the range $0.2 < z < 1.2$. We use the ANNz2 photo-$z$ point estimates to bin our data in the redshift ranges, \ie to select the objects that fall in each redshift bin. We then reconstruct the $N(z)$ by stacking the redshift PDFs of the selected objects for the three codes. \figref{fig:dndz} shows the redshift distributions of the SV data samples compared with their BCC-UFig counterparts. We recall that the BCC-UFig and SV data were subject to the same colour and quality cuts, and restricted to the same portion of the sky: the SPT-E region analysis mask shown in \figref{fig:ufigmaps}. Hence, the inferred redshift distributions should match relatively well since the colours of the BCC-UFig galaxies were shown to correctly match that of the data in \cite{Chang2014bccufig}. This is confirmed by \figref{fig:dndz}: when comparing the left and right panels, the features and relative amplitudes between the $N(z)$ inferred from the three codes are very similar. 

An important difference between the left and right panels is that the {\it true} redshift distributions can be calculated for the BCC-UFig and be compared with the distributions inferred using photometric redshifts.  Analysing the detailed performance of the \photoz codes is beyond the scope of this paper; a full investigation in the context of the weak-lensing SV data samples is presented in \cite{Bonnett2015photozforWL}. However, the results of \figref{fig:dndz} show that most features of data redshift distributions are recovered in the simulation.  For instance, BPZ yields wider $N(z)$s than machine learning methods, but less accurate near $z\sim0.4$ due to the layout of the DES $grizY$ filters and the limitations of the set of template spectra. Also, the redshift distributions inferred by TPZ are narrower than the true underlying distribution. These features persist when selecting galaxies with BPZ or TPZ \photoz point estimates. Selecting with ANNz2 minimises the width of the inferred $N(z)$ from the three methods, and reduces the amount of low-redshift outliers in the third bin.  

\begin{figure}
\hspace*{-2mm}\includegraphics[width=8.7cm, trim = 0.0cm 1.3cm 1.5cm 1.5cm, clip]{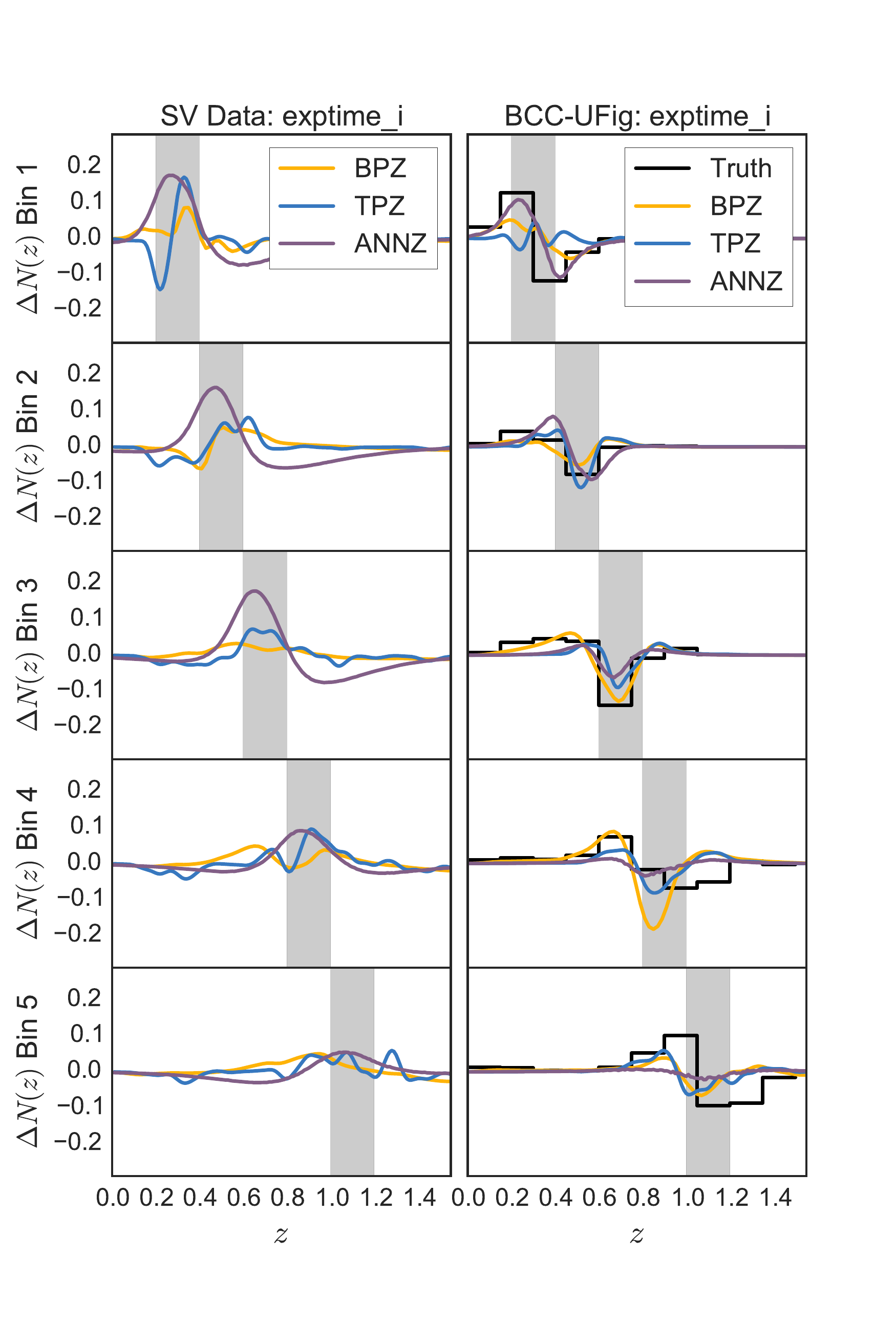}
\caption{Same as \figref{fig:dndzdiff_seeing}, but for {exposure time}. However, note that the fluctuations go in the opposite direction since we compute low minus high values, which corresponds to bad and good regions for exposure time.}
\label{fig:dndzdiff_exp}
\end{figure}

\begin{figure*}\hspace*{-1mm}
\includegraphics[width=18.5cm, trim = 2.5cm 1.6cm 2.3cm 1.4cm, clip]{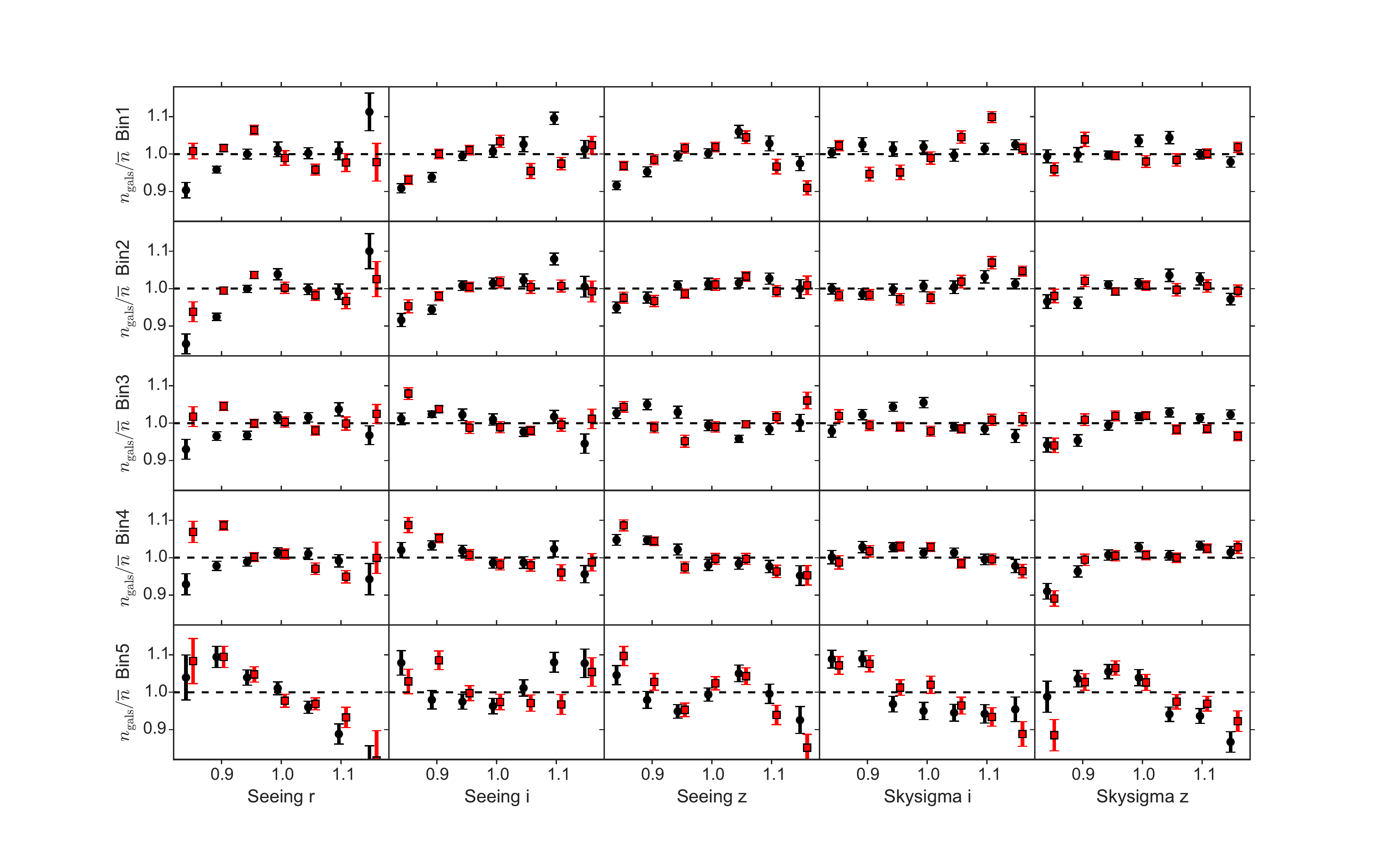}
\caption{{Changes in the galaxy number densities ($n_\mathrm{gals}$) relative to the mean ($\overline{n}$) in each redshift bin, as a function of some observational properties, also normalised to their mean values. Both the SV data (black circles) and BCC-UFig (red squares) exhibit similar fluctuations, which are small but significant compared to the sample variance, calculated using jack-knife re-sampling in 50 sky regions.}}
\label{fig:nulltests}
\end{figure*}

The comparison of true and inferred redshift distributions is not trivial with the SV data given the small sample sizes of spectroscopically confirmed galaxies, especially at high redshift. For this reason, a realistic survey simulation like BCC-UFig is a powerful tool for testing critical analysis stages such as photometric redshift estimation, in regimes that are difficult to explore with the data. More specifically, \figref{fig:dndz} demonstrates that the features seen in the redshift distributions calculated for the SV data are compatible with and well-reproduced by the BCC-UFig simulation.

\bl{We now challenge an assumption made above (and in current SV analyses): the fact that the redshift distributions can be spatially averaged over a large area without accounting for systematics (here the entire SPT-E region)}. While the analysis of SV data is restricted to the most uniform regions, as shown by the analysis mask in \figref{fig:ufigmaps}, these include unavoidable residual depth and quality fluctuations. The mean $N(z)$ is the main quantity of interest for cosmological analyses. However, its variance due to statistical and systematic uncertainties must be evaluated, in order to assess the robustness of theoretical clustering and gravitational lensing predictions using the $N(z)$. {As we will see below, the statistical fluctuations are small, as expected from the area and number density of objects.} However, it is essential to test for residual spatial systematics in the inferred redshift distributions. 

Here we focus on testing the variability of $N(z)$ distributions due to residual depth fluctuations and observational systematics in the SPT-E region. For each of the quantities presented in the previous section (\eg seeing) we compute the median value and use it to split each redshift sample into two subsamples. These cover different regions of the sky, observed under different conditions. We compute the difference between the redshift distributions corresponding to the two patches, \ie taking the redshift distribution of the galaxies in the region where the systematic is above the median value, and subtracting that of the galaxies in the other region. Figures \ref{fig:dndzdiff_seeing} and \ref{fig:dndzdiff_exp} show these differences for the $i$ band seeing and exposure time, respectively; these are two significant sources of spatially-varying depth in the SV data (\eg \citealt{Crocce2015dessvclustering}). \bl{Importantly, the variance obtained by randomly splitting the redshift samples (instead of splitting based on observing conditions) is of the order of $0.01$ in all redshift bins.}

These figures indicate that the $N(z)$ differences are significant compared to the sample variance.  This is expected since good regions (\eg low seeing or high exposure time) have lower noise and better photometry. {As a consequence the \photoz codes will have better overall quality and yield narrower redshift PDFs}. Therefore, the derived redshift distributions when selecting objects in top hat redshift windows will be more accurate. In our difference convention, this translates into a positive bump surrounded by wells in \figref{fig:dndzdiff_seeing} (since we take the difference between low seeing minus high seeing regions), and the opposite in \figref{fig:dndzdiff_exp} (where we compute low exposure time minus high exposure time). This is indeed observed in most bins, even though this depends on the details of the photometric redshift estimation.

While the observed $N(z)$ fluctuations are significant compared to the sample variance, they are small compared to the overall amplitudes shown in \figref{fig:dndz} (less than 5\% in all the cases we tested). This can be seen not only in the histograms of the true redshifts from BCC-UFig, but also in the $N(z)$ inferred from the \photoz codes. In fact, in the right panels corresponding to BCC-UFig, these distributions follow the fluctuations of the true redshifts. This is not the case in all panels, because low and high redshift objects suffer from other issues that make the comparison difficult. In particular, we did not re-weight the redshift distributions to adjust the colour distributions of the training, validation and data samples, as in \cite{Bonnett2015photozforWL, sanchez2014dessva1photoz}. {Such corrections do not affect the comparison between the data and the simulation.} 

This analysis provides an estimate of the order of magnitude of the $N(z)$ fluctuations due to observational sources of systematics and spatially varying depth in the SV data. Provided the mean $N(z)$ is properly characterised, these fluctuations will not bias the cosmological analyses. However, because they are due to residual spatial systematics, they may cause other {types of contamination} in the galaxy catalogues. This is shown in \cite{Crocce2015dessvclustering} and in the next section, where seeing is found to spuriously correlate with the SV data and contaminate the clustering measurements. 

\subsection{Spatial null tests}

We now turn to the spatial properties of the BCC-UFig redshift samples. \figref{fig:nulltests} shows the average galaxy density measured in the previous redshift samples as a function of a few sources of systematics (median exposure time, seeing, and sky sigma). We create these data points by jointly analysing \healpix maps (at $\nside=4096$) of the galaxy redshift bins (SV data and BCC-UFig) and the maps of observing conditions presented in the previous sections. Prior to estimation, all maps are divided by their mean values, so that the observables are dimensionless and concentrated near the central values $(1,1)$ in the panels of \figref{fig:nulltests}. The dynamical range explored by the galaxy densities in each panel depends on the observational quantity under consideration. For example, normalised seeing values are mostly concentrated between $0.9$ and $1.1$, while exposure times span a wider range, as can be verified in \figref{fig:sva1maps}. The error bars are obtained by jack-knife re-sampling in 50 sky regions, which is possible thanks to the large number of objects (greater than $10^4$ in each region).

Analogous galaxy density measurements are shown in \cite{2015Balrog, Crocce2015dessvclustering} using the SV data. \figref{fig:nulltests} shows very similar trends and amplitudes despite using different maps (the median maps instead of the weighted mean maps). The most significant fluctuations are due to the $r$ and $i$ band seeing, particularly in the first and last redshift bins, in agreement with what is found in \cite{2015Balrog, Crocce2015dessvclustering}. Other observational properties create similar but smaller fluctuations.  Remarkably, the BCC-UFig redshift samples exhibit similar galaxy density fluctuations in most bins. \bl{In particular, \figref{fig:nulltests} shows that the characteristic features of the seeing and sky sigma fluctuations as a function of galaxy density are reproduced by BCC-UFig}. This demonstrates that the simulation succeeds in capturing some of the galaxy density fluctuations caused by the systematics considered here. The remaining qualitative and quantitative discrepancies are likely due to the approximations adopted in the simulation. The most significant effect is likely the incorporation of observing conditions at the tile level {instead of the single-epoch images}: the current implementation limits the spatial resolution of systematics to relatively large scales. 

\begin{figure}
\vspace*{3mm}\hspace*{-2mm}\includegraphics[width=8.0cm, trim = 0.3cm 0.2cm 4.0cm 0.0cm, clip]{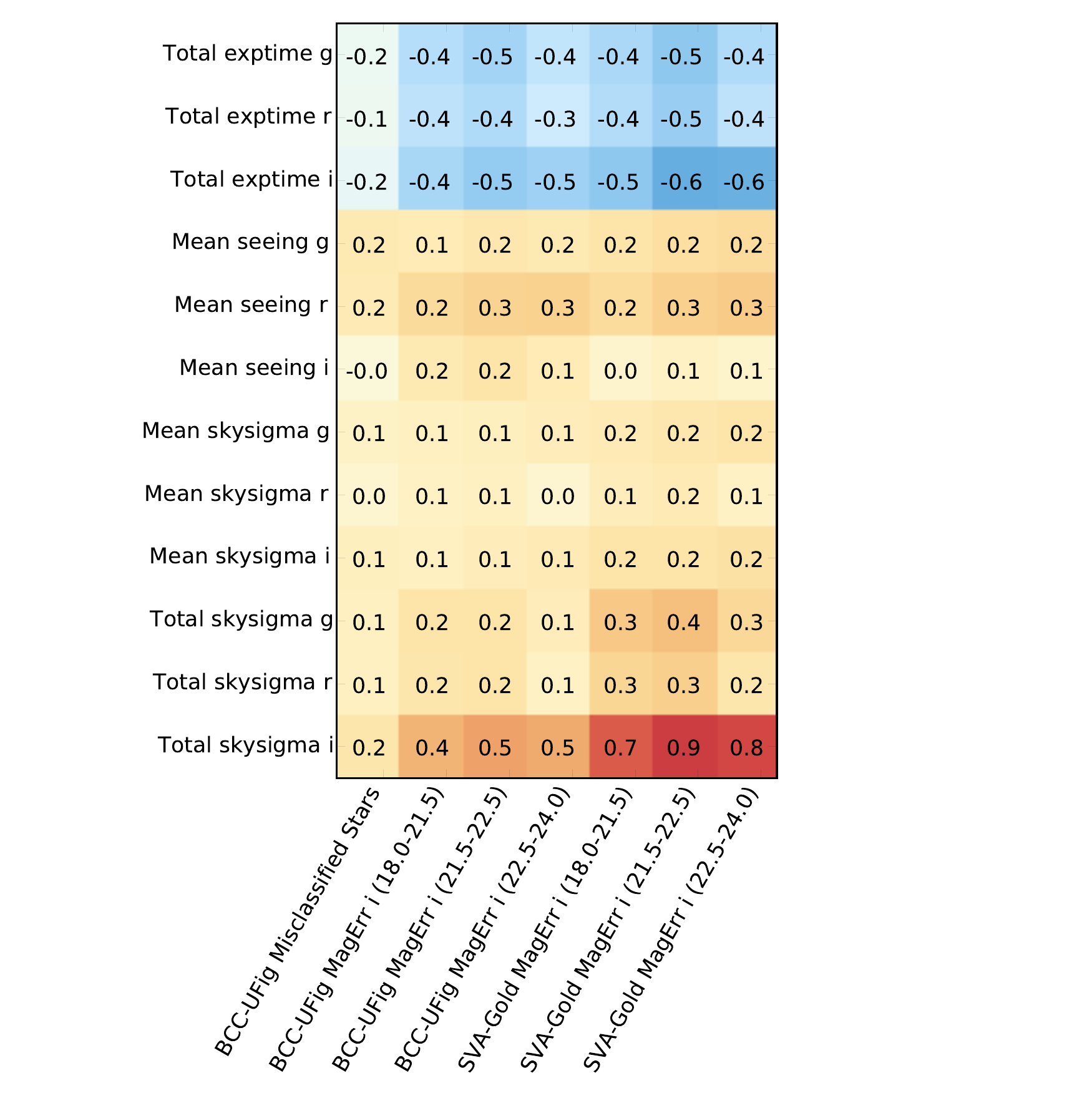}
\caption{{Pearson correlation coefficients between observing conditions, stars missclassified as galaxies in the BCC-UFig reduced data, and mean $i$ band magnitude errors ({\tt magerr\_auto\_}i) in both the BCC-UFig and SVA1-Gold galaxy catalogues (in magnitude bins, using {\tt mag\_auto\_}i). This shows that the maps of the observing conditions are significantly correlated with the stellar contamination and depth fluctuations in the SV data and simulations, therefore capturing the main sources of spatial systematics present in the galaxy samples.}}
\label{fig:syscorr}
\end{figure}

More generally, it is interesting to quantify the extent to which the maps capture depth fluctuations in the data. This is because the effects described above --- spurious spatial variations in the redshift distributions and galaxy densities --- are usually corrected for or marginalised over in cosmological analyses. This is either done at the level of the survey window function or in the measured power spectra or correlation functions. We do not attempt to develop and validate such a model since this must be done in the context of a specific analysis at hand (\eg clustering), which is beyond the scope of this paper. However, we demonstrate that the maps trace the main sources of systematics by showing that they strongly correlate with depth fluctuations and  stellar contamination. \figref{fig:syscorr} shows the Pearson correlation coefficient of some relevant observing condition maps with (1) a map of the stars misclassified as galaxies (by the `modest' classifier) in the BCC-UFig galaxy sample described above; (2) maps of the average $i$ band magnitude errors in the BCC-UFig and SV data (`Gold' catalogue, see \citealt{Crocce2015dessvclustering}) in {\tt mag\_auto\_}i magnitude bins. \figref{fig:syscorr} shows that the exposure time and total sky sigma maps strongly correlate with the magnitude errors in all bands and magnitude bins, in both the data and the simulation, demonstrating that the maps capture most of the depth fluctuations. In fact, a depth map of the SVA1 `Gold' catalogue was constructed using the method described in detail in \cite{Rykoff2015sva1gold}. Briefly, a coarse depth map is first constructed by fitting the magnitude--magnitude error relation of galaxies, exploiting the fact that the magnitude errors satisfy $\sigma_m \propto \sigma_F / F$ where $F$ and $\sigma_F$ are the galaxy flux and its standard deviation. This relation depends on the local limiting magnitude of the survey, which can be estimated in coarse \healpix pixels where there are enough galaxies to obtain precise limiting magnitude estimates (but at low spatial resolution). This map is then refined by constructing a data-driven model of the depth based on the maps of the observing conditions presented here, which are available at very high resolution (using machine learning algorithms, see \citealt{Rykoff2015sva1gold}). The maps were also used in \cite{Crocce2015dessvclustering} to build a linear model of the spurious correlations observed in the angular correlation functions, and correct for them. 

\bl{Note that the correlations between the noise and the magnitude errors are less significant in the simulation than in the data. This is due to the approximation highlighted previously: BCC-UFig is based on simulated coadd images, not on simulated single-epochs. Hence, systematics at scales smaller than the coadds are not resolved. The previous section showed that this approximation yielded correctly reproduced systematics in the galaxy densities and redshift distributions, which are due to large-scale fluctuations of the observing conditions (\eg seeing). However, fluctuations in the noise and depth can be significant on sub-coadd scales. This explains why the correlation coefficient between the map of total sky sigma and the magnitude errors in BCC-UFig is less significant than what found in the data.} 

\bl{
Finally, as shown in \figref{fig:syscorr}, the observing condition maps correlate with the stellar contamination in the BCC-UFig samples. This test cannot be performed with the SV data since object types are only available for a small sample of  spectroscopically confirmed sources, restricted to a small region of the sky, as shown in the previous sections. A large, realistic simulation such as BCC-UFig allows us to confirm that the observing conditions trace the main sources of spatial systematics in the galaxy samples, and can be used to model and remove them in cosmological studies. 
}

\section{Conclusions and outlook}

We detailed a method to extract and project the properties of multi-epoch galaxy surveys onto the sky, making use of the properties of the images and the \healpix pixelisation. We applied this technique to the DES SV data and mapped the main sources of observational systematics, including the average properties of seeing, airmass, and sky sigma. These maps will be made publicly available in the forthcoming DES data releases, {and are currently used in analyses of SV data \citep[\eg][]{2015arXiv150403002V, Crocce2015dessvclustering, Jarvis2015svshear, Becker2015svshear, Giannantonio2015cmbxlssdessv}.}

{High-resolution maps of the observing conditions can be used as templates to identify, model, and mitigate spatial systematics or residual contamination in the data.} As an illustration, we measured the galaxy densities and redshift distributions of DES SV tomographic redshift galaxy samples, and showed that they were significantly affected by the observational conditions of the survey due to residual depth and photometry fluctuations. These systematics are correctly mitigated in current SV analyses thanks to the sky masks and corrections to the two-point correlation measurements validated by stringent null-tests \citep[see \eg][]{Giannantonio2015cmbxlssdessv, Becker2015svshear, Crocce2015dessvclustering}. However, it will be increasingly difficult to keep them under control in future studies. For instance, restrictive sky masks can remove unreliable regions but often discard hard-won data and do not alleviate the need to treat spatial systematics in the retained regions. As the depth and sensitivity of the survey increase, these systematics will become increasingly significant compared to statistical errors. The power of masking or current correction techniques is also limited since they rely on templates and contamination models which are not validated against simulations. 

One approach to resolve these issues, \ie assess the significance of systematics and validate the techniques to mitigate them, is to resort to realistic image simulations. All the tests and analyses of this paper were performed in parallel on galaxy samples obtained by processing the BCC-UFig in the same way as the SV data and applying the same quality and selection cuts. These simulated galaxy samples include spatial systematics since the image simulations incorporate the actual SV observing conditions. \bl{Even with the approximation of simulating coadd images instead of single-epochs, we found that the principal effects of spatial systematics observed in the galaxy densities and redshift distributions were successfully reproduced by the BCC-UFig galaxy samples}. Furthermore, the data and the simulation agreed quantitatively in many cases, showing that the current BCC-UFig simulation, even with known limitations, is sufficiently realistic to study a range of effects. The availability of the ground truth in the simulation (\eg the true redshifts) allowed us to quantify the significance of the systematic density and redshift fluctuations for the first time, {and to demonstrate that the observing condition maps capture systematics such as depth fluctuations and stellar contamination}. Pursuing this route will be essential for the future DES studies, since these fluctuations will have to be carefully characterised and mitigated. Future versions of the BCC-UFig simulation will be more realistic and reproduce spatial systematics at higher resolution. Combining them with high-resolution maps of the observing conditions and the effective transfer function measured by \textsc{Balrog} \citep{2015Balrog} will allow us to fully exploit the potential of DES data for cosmological studies. These complementary avenues will be essential to correctly interpret the deep, high-cadence data delivered by the Large Synoptic Survey Telescope (LSST), where both the statistical power and the impact of the observing conditions will be increased by many orders of magnitude \citep[\eg][]{2009arXiv0912.0201L, 2012arXiv1211.0310L, 2011PASP..123..596J, 2014SPIE.9149E..0CC}.

\section{Acknowledgements}

BL, HVP and FE are supported by STFC and the European Research Council under the European Community's Seventh Framework Programme (FP7/2007- 2013) / ERC grant agreement no 306478-CosmicDawn. This work was supported in part by National Science Foundation Grant No. PHYS-1066293 and the hospitality of the Aspen Center for Physics. CC, AR, AA and CB are supported by in part the Swiss National Foundation grants 200021-149442 and 200021-143906. FS acknowledges financial support provided by CAPES under contract No. 3171-13-2. ML is partially supported by FAPESP and CNPq. We acknowledge use of the \healpix software package \citep{healpix1}. 

We are grateful for the extraordinary contributions of our CTIO colleagues and the DES Camera, Commissioning and Science Verification teams in achieving the excellent instrument and telescope conditions that have made this work possible. The success of this project also relies critically on the expertise and dedication of the DES Data Management organization. 

This paper has gone through internal review by the DES collaboration. The DES and Fermilab publication numbers are DES-2015-0098 and FERMILAB-PUB-15-310-A-AE, respectively.

Funding for the DES Projects has been provided by the U.S. Department of Energy, the U.S. National Science Foundation, the Ministry of Science and Education of Spain, 
the Science and Technology Facilities Council of the United Kingdom, the Higher Education Funding Council for England, the National Center for Supercomputing 
Applications at the University of Illinois at Urbana-Champaign, the Kavli Institute of Cosmological Physics at the University of Chicago, 
the Center for Cosmology and Astro-Particle Physics at the Ohio State University,
the Mitchell Institute for Fundamental Physics and Astronomy at Texas A\&M University, Financiadora de Estudos e Projetos, 
Funda{\c c}{\~a}o Carlos Chagas Filho de Amparo {\`a} Pesquisa do Estado do Rio de Janeiro, Conselho Nacional de Desenvolvimento Cient{\'i}fico e Tecnol{\'o}gico and 
the Minist{\'e}rio da Ci{\^e}ncia, Tecnologia e Inova{\c c}{\~a}o, the Deutsche Forschungsgemeinschaft and the Collaborating Institutions in the Dark Energy Survey. 
The DES data management system is supported by the National Science Foundation under Grant Number AST-1138766.

The Collaborating Institutions are Argonne National Laboratory, the University of California at Santa Cruz, the University of Cambridge, Centro de Investigaciones En{\'e}rgeticas, 
Medioambientales y Tecnol{\'o}gicas-Madrid, the University of Chicago, University College London, the DES-Brazil Consortium, the University of Edinburgh, 
the Eidgen{\"o}ssische Technische Hochschule (ETH) Z{\"u}rich, 
Fermi National Accelerator Laboratory, the University of Illinois at Urbana-Champaign, the Institut de Ci{\`e}ncies de l'Espai (IEEC/CSIC), 
the Institut de F{\'i}sica d'Altes Energies, Lawrence Berkeley National Laboratory, the Ludwig-Maximilians Universit{\"a}t M{\"u}nchen and the associated Excellence Cluster Universe, 
the University of Michigan, the National Optical Astronomy Observatory, the University of Nottingham, The Ohio State University, the University of Pennsylvania, the University of Portsmouth, 
SLAC National Accelerator Laboratory, Stanford University, the University of Sussex, and Texas A\&M University.

The DES participants from Spanish institutions are partially supported by MINECO under grants AYA2012-39559, ESP2013-48274, FPA2013-47986, and Centro de Excelencia Severo Ochoa SEV-2012-0234.
Research leading to these results has received funding from the European Research Council under the European Union's Seventh Framework Programme (FP7/2007-2013) including ERC grant agreements 
 240672, 291329, and 306478.

\footnotesize{
  \bibliographystyle{mn2e_eprint}
\providecommand{\eprint}[1]{\href{http://arxiv.org/abs/#1}{arXiv:#1}}	
  \bibliography{bib}
}
\normalsize

\section*{Affiliations}
 \noindent
$^{1}$ Department of Physics \& Astronomy, University College London, Gower Street, London, WC1E 6BT, UK\\ 
$^{2}$ Department of Physics, ETH Zurich, Wolfgang-Pauli-Strasse 16, CH-8093 Zurich, Switzerland\\ 
$^{3}$ Institut de Ci\`encies de l'Espai, IEEC-CSIC, Campus UAB, Carrer de Can Magrans, s/n,  08193 Bellaterra, Barcelona, Spain\\ 
$^{4}$ Department of Physics, Stanford University, 382 Via Pueblo Mall, Stanford, CA 94305, USA\\ 
$^{5}$ Kavli Institute for Particle Astrophysics \& Cosmology, P. O. Box 2450, Stanford University, Stanford, CA 94305, USA\\ 
$^{6}$ Institut de F\'{\i}sica d'Altes Energies, Universitat Aut\`onoma de Barcelona, E-08193 Bellaterra, Barcelona, Spain\\ 
$^{7}$ SLAC National Accelerator Laboratory, Menlo Park, CA 94025, USA\\ 
$^{8}$ Department of Astronomy, University of Illinois, 1002 W. Green Street, Urbana, IL 61801, USA\\ 
$^{9}$ National Center for Supercomputing Applications, 1205 West Clark St., Urbana, IL 61801, USA\\ 
$^{10}$ Laborat\'orio Interinstitucional de e-Astronomia - LIneA, Rua Gal. Jos\'e Cristino 77, Rio de Janeiro, RJ - 20921-400, Brazil\\ 
$^{11}$ Observat\'orio Nacional, Rua Gal. Jos\'e Cristino 77, Rio de Janeiro, RJ - 20921-400, Brazil\\ 
$^{12}$ Center for Cosmology and Astro-Particle Physics, The Ohio State University, Columbus, OH 43210, USA\\ 
$^{13}$ Department of Physics, The Ohio State University, Columbus, OH 43210, USA\\ 
$^{14}$ Institute of Cosmology \& Gravitation, University of Portsmouth, Portsmouth, PO1 3FX, UK\\ 
$^{15}$ Department of Physics, University of Arizona, Tucson, AZ 85721, USA\\ 
$^{16}$ Centro de Investigaciones Energ\'eticas, Medioambientales y Tecnol\'ogicas (CIEMAT), Madrid, Spain\\ 
$^{17}$ Fermi National Accelerator Laboratory, P. O. Box 500, Batavia, IL 60510, USA\\ 
$^{18}$ Institute of Astronomy, University of Cambridge, Madingley Road, Cambridge CB3 0HA, UK\\ 
$^{19}$ Kavli Institute for Cosmology, University of Cambridge, Madingley Road, Cambridge CB3 0HA, UK\\ 
$^{20}$ Department of Physics and Astronomy, University of Pennsylvania, Philadelphia, PA 19104, USA\\ 
$^{21}$ Carnegie Observatories, 813 Santa Barbara St., Pasadena, CA 91101, USA\\ 
$^{22}$ CNRS, UMR 7095, Institut d'Astrophysique de Paris, F-75014, Paris, France\\ 
$^{23}$ Sorbonne Universit\'es, UPMC Univ Paris 06, UMR 7095, Institut d'Astrophysique de Paris, F-75014, Paris, France\\ 
$^{24}$ Jodrell Bank Center for Astrophysics, School of Physics and Astronomy, University of Manchester, Oxford Road, Manchester, M13 9PL, UK\\ 
$^{25}$ George P. and Cynthia Woods Mitchell Institute for Fundamental Physics and Astronomy, and Department of Physics and Astronomy, Texas A\&M University, College Station, TX 77843,  USA\\ 
$^{26}$ Department of Physics, Ludwig-Maximilians-Universit\"at, Scheinerstr. 1, 81679 M\"unchen, Germany\\ 
$^{27}$ Excellence Cluster Universe, Boltzmannstr.\ 2, 85748 Garching, Germany\\ 
$^{28}$ Jet Propulsion Laboratory, California Institute of Technology, 4800 Oak Grove Dr., Pasadena, CA 91109, USA\\ 
$^{29}$ Department of Astronomy, University of Michigan, Ann Arbor, MI 48109, USA\\ 
$^{30}$ Department of Physics, University of Michigan, Ann Arbor, MI 48109, USA\\ 
$^{31}$ Kavli Institute for Cosmological Physics, University of Chicago, Chicago, IL 60637, USA\\ 
$^{32}$ Max Planck Institute for Extraterrestrial Physics, Giessenbachstrasse, 85748 Garching, Germany\\ 
$^{33}$ Universit\"ats-Sternwarte, Fakult\"at f\"ur Physik, Ludwig-Maximilians Universit\"at M\"unchen, Scheinerstr. 1, 81679 M\"unchen, Germany\\ 
$^{34}$ Cerro Tololo Inter-American Observatory, National Optical Astronomy Observatory, Casilla 603, La Serena, Chile\\ 
$^{35}$ Australian Astronomical Observatory, North Ryde, NSW 2113, Australia\\ 
$^{36}$ Departamento de F\'{\i}sica Matem\'atica,  Instituto de F\'{\i}sica, Universidade de S\~ao Paulo,  CP 66318, CEP 05314-970 S\~ao Paulo, Brazil\\ 
$^{37}$ Department of Astronomy, The Ohio State University, Columbus, OH 43210, USA\\ 
$^{38}$ Instituci\'o Catalana de Recerca i Estudis Avan\c{c}ats, E-08010 Barcelona, Spain\\ 
$^{39}$ Department of Physics and Astronomy, Pevensey Building, University of Sussex, Brighton, BN1 9QH, UK\\ 
$^{40}$ Instituto de F\'\i sica, UFRGS, Caixa Postal 15051, Porto Alegre, RS - 91501-970, Brazil\\ 
$^{41}$ Department of Physics, University of Illinois, 1110 W. Green St., Urbana, IL 61801, USA\\ 
$^{42}$ Argonne National Laboratory, 9700 South Cass Avenue, Lemont, IL 60439, USA\\ 
$^{43}$SEPnet, South East Physics Network, (www.sepnet.ac.uk)\\


\end{document}